\documentclass[12pt,number]{elsarticle}
\usepackage{a4wide,amsmath,amssymb,epsfig}
\usepackage[tight]{subfigure}

\newcommand{\D}{\mbox{$\,\mathrm{d}$}}
\newcommand{\Dx}{\mbox{$\mathrm{d}$}}

\newcommand{\Pcl}{\mbox{$P_l^\mathrm{C}$}}
\newcommand{\Pco}{\mbox{$P_o^\mathrm{C}$}}

\newcommand{\mm}{\mbox{$\mathrm{mm}$}}
\newcommand{\kN}{\mbox{$\mathrm{kN}$}}

\newcommand{\e}[1]{\ensuremath{\times 10^{#1}}}

\journal{Thin-Walled Structures}

\begin{document}

\title{Cellular buckling in I-section struts}

\author{M. Ahmer Wadee\corref{cor1}}
\ead{a.wadee@imperial.ac.uk}

\author{Li Bai}
\ead{li.bai06@imperial.ac.uk}

\cortext[cor1]{Corresponding author}

\address{Department of Civil and Environmental Engineering, Imperial
  College London, South Kensington Campus, London SW7 2AZ, UK}


\begin{frontmatter}
  
\begin{abstract}

  An analytical model that describes the interactive buckling of a
  thin-walled I-section strut under pure compression based on
  variational principles is presented. A formulation combining the
  Rayleigh--Ritz method and continuous displacement functions is used
  to derive a system of differential and integral equilibrium
  equations for the structural component. Numerical continuation
  reveals progressive cellular buckling (or \emph{snaking}) arising
  from the nonlinear interaction between the weakly stable global
  buckling mode and the strongly stable local buckling mode. The
  resulting behaviour is highly unstable and when the model is
  extended to include geometric imperfections it compares excellently
  with some recently published experiments.

\end{abstract}

\end{frontmatter}

\section{Introduction}
\label{sec:intro}

The buckling of struts and columns represents the most common type of
structural instability problem \cite{TG61}. However, when the
compression member is made from slender metallic plate elements they
are well known to suffer from a variety of different elastic
instability phenomena. In the current work, the classic problem of a
strut under axial compression made from a linear elastic material with
an open and doubly-symmetric cross-section -- an ``I-section''
\cite{Neut1969,Hancock1981} -- is studied in detail using an
analytical approach.  Under this type of loading, long members are
primarily susceptible to a global (or overall) mode of instability
namely Euler buckling, where flexure about the weak axis occurs once
the theoretical Euler buckling load is reached. However, when the
individual plate elements of the strut cross-section, namely the
flanges and the web, are relatively thin or slender, elastic local
buckling of these may also occur; if this happens in combination with
the global instability, then the resulting behaviour is usually far
more unstable than when the modes are triggered individually. Recent
work on the interactive buckling of struts include experimental and
finite element studies \cite{Becque2009expt,Becque2009num}, where the
focus was on the behaviour of struts made from stainless
steel. However, the more generic finding that the members had an
increased sensitivity to imperfections was highlighted. Other
structural components that are known to suffer from the interaction of
local and global instability modes are thin-walled beams under uniform
bending \cite{WG2012}, sandwich struts \cite{HW1998,ijnm_rbt},
stringer-stiffened and corrugated plates
\cite{Koiter1976,Pignataro_TWS2000} and built-up, compound or
reticulated columns \cite{TH1973}.

Apart from the aforementioned work where some numerical modelling was
presented \cite{Becque2009num}, the formulation of a mathematical
model accounting for the nonlinear interactive buckling behaviour has
not been forthcoming. The current work presents the development of a
variational model that accounts for the mode interaction between
global Euler buckling and local buckling of a flange such that the
perfect and imperfect elastic post-buckling response of the strut can
be evaluated. A system of nonlinear ordinary differential equations
subject to integral constraints is derived and is solved using the
numerical continuation package \textsc{Auto} \cite{Auto2007}. It is
indeed found that the system is highly unstable when interactive
buckling is triggered; snap-backs in the response, showing sequential
destabilization and restabilization and a progressive spreading of the
initial localized buckling mode, are also revealed. This latter type of
response has become known in the literature as \emph{cellular
  buckling} \cite{Hunt2000} or \emph{snaking} \cite{BurkeKnobloch2007}
and it is shown to appear naturally in the numerical results of the
current model. As far as the authors are aware, this is the first time
this phenomenon has been captured analytically in struts undergoing
Euler and local buckling simultaneously. Similar behaviour has been
discovered in various other mechanical systems such as in the
post-buckling of cylindrical shells \cite{Hunt2003}, the sequential
folding of geological layers \cite{MAW_jmps05} and most recently in
the lateral buckling of thin-walled beams under pure bending
\cite{WG2012}.

Experimental results from the literature
\cite{Becque2009expt,Becque_thesis} are used for validation
purposes. The mechanical destabilization and the nature of the
post-buckling deformation compare excellently with the current
model. This demonstrates that the fundamental physics of this system
is captured by the analytical approach both qualitatively and
quantitatively. A brief discussion is presented on how the current
model could be enhanced and then conclusions are drawn.

\section{Analytical Model}

\begin{figure}[htb]
\centerline{\psfig{figure=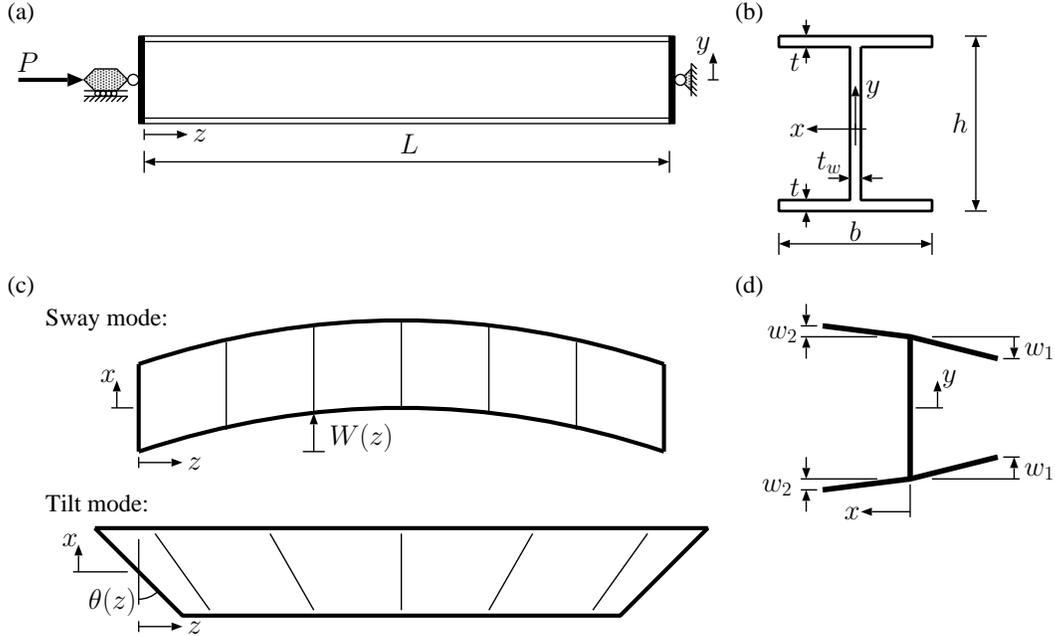,width=140mm}}
\caption{(a) Elevation of an I-section strut of length $L$ that is
  compressed axially by a force $P$. The lateral and longitudinal
  coordinates are $y$ and $z$ respectively. (b) Cross-section of
  strut; the transverse coordinate is $x$. (c) Sway and tilt
  components of the minor axis global buckling mode. (d) Local
  buckling mode: out-of-plane flange displacement functions
  $w_i(x,z)$; note the linear distribution in the $x$ direction.}
\label{fig:strutswaylocal}
\end{figure}

Consider a thin-walled I-section strut of length $L$ made from a
linear elastic, homogeneous and isotropic material with Young's
modulus $E$ and Poisson's ratio $\nu$. It is loaded by an axial force
$P$ that is applied at the centroid of the cross-section, as shown in
Figure \ref{fig:strutswaylocal}(a) and (b) respectively,
with rigid end plates that transfer the force uniformly to the entire
cross-section.  The web is assumed to provide a simple support to both
flanges and not to buckle locally under the axial compression, an
assumption that is justified later. In the current study, the total
cross-section depth is $h$ with each flange having width $b$ and
thickness $t$. It is assumed currently that the I-section is
effectively made up from two channel members connected back-to-back;
hence, the assumption is that the web thickness $t_w=2t$, a type of
arrangement that has been used in recent experimental studies
\cite{Becque2009expt,WG2012,Becque_thesis}. The strut length $L$ is
varied such that in one case, which is presented later, Euler buckling
about the weaker $y$-axis occurs before any flange buckles locally and
in the other case the reverse is true -- flange local buckling is
critical.

The formulation begins with the definitions for both the global and
the local modal displacements. Timoshenko beam theory is assumed,
meaning that the effect of shear is not neglected as in standard
Euler--Bernoulli beam theory. Although it turns out that the effect of
shear is only minor, it is necessary to account for it since it
provides the key terms within the total potential energy that allow
buckling mode interaction to be modelled \cite{WG2012,HW1998}. To
account for shear, two generalized coordinates $q_s$ and $q_t$,
defined as the amplitudes of the degrees of freedom known as ``sway''
and ``tilt'' \cite{HW1998} are introduced to model the global mode, as
shown in Figure \ref{fig:strutswaylocal}(c), where the lateral
displacement $W$ and the rotation $\theta$ are given by the following
expressions:
\begin{equation}
  W(z) = q_s L \sin\frac{\pi z}{L}, \quad
  \theta(z) = q_t \pi \cos\frac{\pi z}{L}.
 \label{eq:swaytilt}
\end{equation}
For the present case, the shear strain in the $xz$ plane,
$\gamma_{xz}$, is included and is given by the following expression:
\begin{equation}
  \gamma_{xz}=\frac{\D W}{\D z}-\theta=\left(q_s-q_t\right)\pi
  \cos\frac{\pi z}{L}.
\end{equation}
Of course, Euler--Bernoulli beam theory would imply that since
$\gamma_{xz}=0$, then $q_s = q_t$.

\begin{figure}[htb]
  \centering
  \psfig{figure=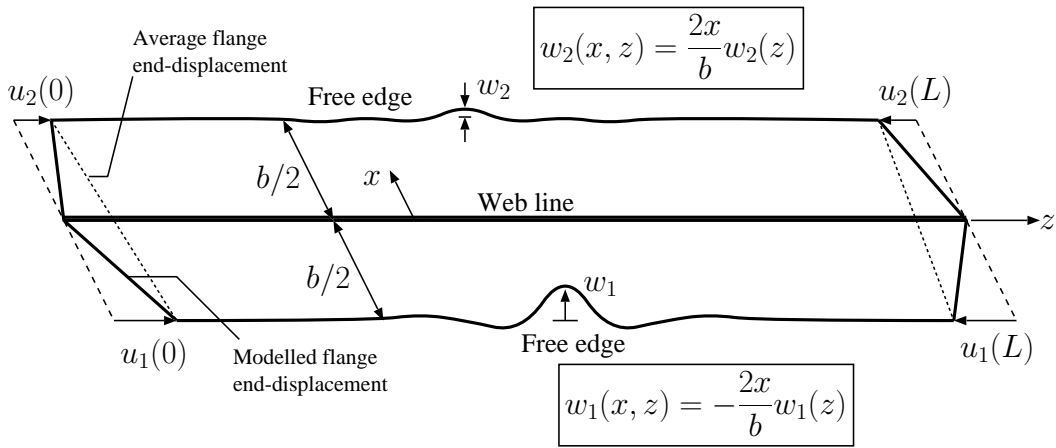,width=140mm}
  \caption{Displacement functions of local buckling mode in
    flanges. Longitudinal and lateral flange displacements $u_i(x,z)$
    and $w_i(x,z)$ respectively. Note the linear distributions in $x$
    direction and the average end-displacement, as opposed to the
    modelled flange end-displacement, which is used to calculate the
    local contribution to the work done.}
\label{fig:flange_uw}
\end{figure}

The local mode is modelled with appropriate boundary
conditions. Moreover, the possibility of a distinct local buckling
mode occurring before global buckling implies that the entire flange
may deflect. However, if the interaction between local and global
modes occurs then the symmetry of the local buckling mode would be
broken and the flanges would not buckle with the same
displacement. Hence, two separate lateral displacement functions $w_1$
and $w_2$ need to be defined, as shown in Figure
\ref{fig:strutswaylocal}(d), to allow for the break in symmetry. Since
the outstands of the flanges have free edges, whereas the web is
assumed to provide no more than a simple support to the flanges, a
linear distribution is assumed in the $x$ direction; Bulson
\cite{Bulson} showed this distribution is correct for the local
buckling eigenmode for that type of rectangular plate.  For the local
mode in-plane displacements $u_i$, the distributions are also assumed
to be linear in $x$, as shown in Figure \ref{fig:flange_uw}.
This is in fact another consequence of the Timoshenko beam theory
assumption where plane sections are assumed to remain plane. These
assumptions lead to the following expressions for the local
out-of-plane displacements $w_i$ with the in-plane displacements $u_i$:
\begin{equation}
 w_i(x,z) =(-1)^{i} \left( \frac{2x}{b} \right)w_i(z), \quad u_i(x,z)
 =(-1)^{i} \left(\frac{2x}{b}\right) u_i(z),
\end{equation}
where $i=\{1,2\}$ throughout the current article. The transverse
in-plane displacement $v(z,x)$ is assumed to be small and is hence
neglected for the current case; this reflects the findings from Koiter
and Pignataro \cite{Koiter1976} for rectangular plates with three
pinned edges and one free edge.

Since, in practice, perfect geometries do not exist, an initial
out-of-straightness in the $x$-direction, $W_0$, is introduced as a
global imperfection to the web and flanges in the current model. An
initial rotation of the plane section $\theta_0$ is also introduced to
simulate the out-of-straightness in the flanges. The expressions for
$W_0$ and $\theta_0$ are given by:
\begin{equation}
W_0 = q_{s0} L \sin\frac{\pi z}{L}, \quad
  \theta_0 = q_{t0} \pi \cos\frac{\pi z}{L},
\end{equation}
and are analogous to Equation (\ref{eq:swaytilt}).  Note that the
assumption of Timoshenko beam theory implies that shear strains in the
$xz$ plane due to the initial imperfection are also introduced.

\subsection{Total potential energy}

\begin{figure}[htbp]
  \centering
  \subfigure[]{\psfig{figure=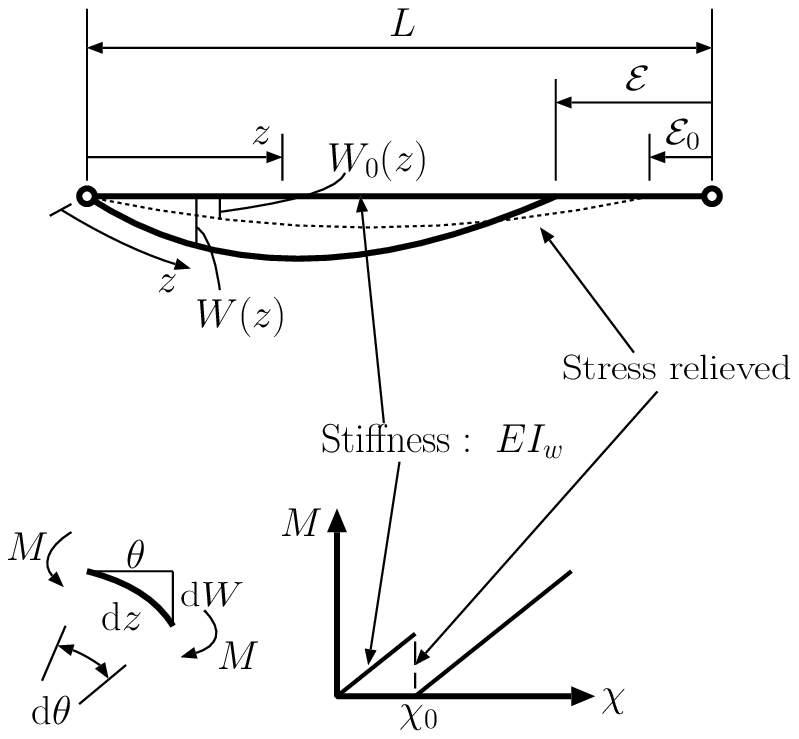,width=90mm}}\quad
  \subfigure[]{\psfig{figure=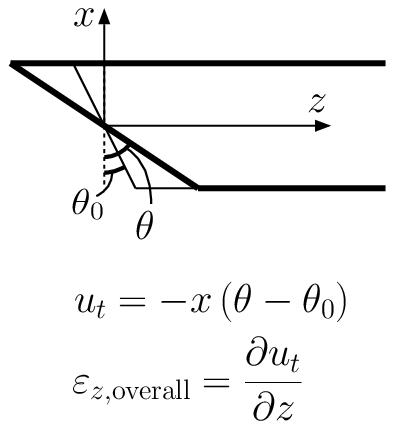,width=50mm}}
  \caption{Introduction of geometric imperfections $W_0$ and
    $\theta_0$ in (a) the web and (b) the flanges.}
  \label{fig:impf}
\end{figure}

The total potential energy, $V$, was determined with the main
contributions being the global and local bending energy $U_{bo}$ and
$U_{bl}$ respectively, the membrane energy $U_m$, and the work done
$P\mathcal{E}$. Note that the global bending energy $U_{bo}$ only
comprises the bending energy stored in the web, since the membrane
energy stored in the flanges accounts for the effect of bending in the
flanges through the tilt mode. The initial out-of-straightness
$W_0(z)$ is stress-relieved \cite{TH1984,Wadee2000}, implying that the
elemental moment $M$ drops to zero as illustrated in Figure
\ref{fig:impf}(a).
The global bending energy involves the second derivative of $W$ and
$W_0$ and is hence given by:
\begin{equation}
  U_{bo} =\frac12 EI_w \int_0^L \left(\ddot{W}-\ddot{W_0} \right)^2 \D
  z =\frac12 EI_w \int_0^L \left(q_s-q_{s0}\right)^2 \frac{\pi^4}{L^2}
  \sin^2\frac{\pi z}{L} \D z,
\end{equation}
where dots represent differentiation with respect to $z$ and
$I_w=t_w^3(h-2t)/12$ is the second moment of area of the web about the
global weak axis. Obviously, for the case where $t_w=2t$, the
expression becomes $I_w=2t^3(h-2t)/3$. The local bending energy,
accounting for both flanges, is determined as:
\begin{equation}
\begin{aligned}
  U_{bl} & = D \int_0^L \left[ \int_{-b/2}^0 B_1 \Dx x +
    \int_0^{b/2} B_2 \Dx x \right] \Dx z \\
  & = D \int_0^L \left[ \frac{b}{6}
    \left(\ddot{w}_1^2+\ddot{w}_2^2\right) + \frac{4\left(1-\nu
    \right)}{b}  \left(\dot{w}_1^2+\dot{w}_2^2 \right)\right] \Dx z,
\end{aligned}
\end{equation}   
where $B_i$, the contribution from $w_i$ to the standard expression
for the incremental strain energy from bending a plate \cite{TG61}, is
given by:
\begin{equation}
  B_i = \left(\frac{\partial^2 w_i}{\partial z^2}+\frac{\partial^2
      w_i}{\partial x^2} \right)^2 - 2\left(1-\nu \right)
  \left[\frac{\partial^2 w_i}{\partial z^2} \frac{\partial^2
      w_i}{\partial x^2} -\left(\frac{\partial^2 w_i}{\partial
        z \partial x} \right)^2 \right],
\end{equation}
with $D=Et^3/[12(1-\nu^2)]$ being the plate flexural rigidity. The
buckled configuration of the flange plate involves double curvature in
the $x$ and $z$ directions, indicating the non-developable nature of
plate deformation. The so-called membrane strain energy ($U_m$) is
derived from considering the direct strains ($\varepsilon$) and the
shear strains ($\gamma$) in the flanges thus:
\begin{equation}
    U_m = U_d + U_s = \int_0^L \left[ \int_{-h/2}^{-h/2+t} F \D y +
      \int_{h/2-t}^{h/2} F \D y \right] \Dx z,
\end{equation}
where:
\begin{equation}
  \begin{split}
    F = \frac12 & \biggl\{ \int_{-b/2}^{0} \left[ E \left(
        \varepsilon^2_{z1} +\varepsilon^2_{x1} + 2 \nu
        \varepsilon_{z1} \varepsilon_{x1} \right) + G \gamma^2_{xz1}
    \right] \Dx x\\
    & \quad + \int_{0}^{b/2} \left[ E \left(
        \varepsilon^2_{z2} +\varepsilon^2_{x2} + 2 \nu
        \varepsilon_{z2}\varepsilon_{x2} \right) + G \gamma^2_{xz2}
    \right] \Dx x \biggl\}.
  \end{split}
\end{equation}
The transverse component of strain $\varepsilon_{xi}$ is neglected
since it has been shown that it has no effect on the post-buckling
behaviour of a long plate with three simply-supported edges and one
free edge \cite{Koiter1976}. The longitudinal strain $\varepsilon_z$
has to be modelled separately for different outstand flanges. Recall
that the tilt component of the in-plane displacement from the global
mode, including the initial imperfection, is given by $u_t =
-(\theta-\theta_0) x$ as shown in Figure \ref{fig:impf}(b); hence:
\begin{equation}
  \varepsilon_{z,\mathrm{global}} = \frac{\partial u_t}{\partial z} =
  x \left(q_t-q_{t0}\right) \frac{\pi^2}{L} \sin\frac{\pi z}{L}.
\end{equation}
The local mode contribution is based on von K\'arm\'an plate theory. A
pure in-plane compressive strain $\Delta$ is also included. The direct
strains in the compression and tension side of the flanges, denoted as
$\varepsilon_{z1}$ and $\varepsilon_{z2}$ respectively, are given by
the general expression:
\begin{equation}
\begin{aligned}
  \varepsilon_{zi} &= \varepsilon_{z,\mathrm{global}} - \Delta +
  \frac{\partial u_i}{\partial z}
  + \frac12 \left(\frac{\partial w_i}{\partial z} \right)^2\\
  &= x \left(q_t-q_{t0}\right) \frac{\pi^2}{L} \sin \frac{\pi z}{L} -
  \Delta + (-1)^i \left(\frac{2x}{b}\right) \dot{u_i} +
  \frac{2x^2}{b^2} \dot{w}_i^2.
\end{aligned}
\end{equation}
The strain energy from direct strains ($U_d$) is thus, assuming that
$h \gg t$:
\begin{equation}
\begin{aligned}
  U_d = Etb \int_0^L & \biggl\{ \frac{b^2}{12}
  \left(q_t-q_{t0}\right)^2 \frac{\pi^4}{L^2} \sin^2 \frac{\pi z}{L} +
  \Delta^2 +\frac16 \left({\dot{u}_1}^2+{\dot{u}_2}^2 \right)
  +\frac{1}{40}
  \left({\dot{w}_1}^4+{\dot{w}_2}^4 \right) \\
  & - \left(q_t-q_{t0}\right) \frac{b\pi^2}{2L} \sin \frac{\pi z}{L}
  \left[ \frac13 \left({\dot{u}}_1-{\dot{u}}_2\right)+\frac18
    \left({\dot{w}_1}^2-{\dot{w}_1}^2 \right)\right] - \frac12 \Delta
  \left(\dot{u}_1+\dot{u}_2\right) \\
  & -\frac16 \Delta \left({\dot{w}_1}^2+{\dot{w}_2}^2\right) +\frac18
  \left(\dot{u}_1 {\dot{w}_1}^2+ \dot{u}_2 {\dot{w}_2}^2
  \right)+\frac{h}{b}\Delta^2 \biggr\} \D z,
\end{aligned}
\end{equation} 
where, apart from the final term which represents the energy stored in
the web, the contributions are from the direct strains in both
flanges.  The shear strain energy $U_s$ contains the shear modulus
$G$, which is given by $E/[2(1+\nu)]$ for a homogeneous and isotropic
material. The shear strain $\gamma_{xz}$ contributions are also
modelled separately for the compression and the tension side of the
flanges. The expression for each outstand is given by the general
expression:
\begin{equation}
\begin{aligned}
  \gamma_{xzi} &= \frac{\partial}{\partial z} \left(W - W_0\right) -
  \left(\theta-\theta_0\right) + \frac{\partial u_i}{\partial
    x}+\frac{\partial w_i}{\partial
    z}\frac{\partial w_i}{\partial x} \\
  &= \left(q_s - q_t-q_{s0}+q_{t0}\right) \pi \cos \frac{\pi z}{L} +
  (-1)^i \left( \frac{2}{b} \right) u_i +\frac{4x}{b^2} w_i \dot{w_i}.
\end{aligned}
\end{equation}
The expression for the strain energy from shear is thus:
\begin{equation}
 \begin{aligned}
   U_s = Gtb \int_0^L & \biggl[\left(q_s - q_t-q_{s0}+q_{t0}\right)^2
   \pi^2 \cos^2 \frac{\pi z}{L} \\
   & - \left(q_s - q_t-q_{s0}+q_{t0}\right) \frac{\pi}{b} \cos
   \frac{\pi z}{L} \left(2u_1-2u_2 +w_1 \dot{w}_1-w_2 \dot{w}_2 \right) \\
   & + \frac{2}{b^2} \left( {u_1}^2 +{u_2}^2 + \frac13 {w_1}^2
     {\dot{w}_1}^2 + \frac13 {w_2}^2 {\dot{w}_2}^2 + u_1 w_1 \dot{w}_1
     + u_2 w_2 \dot{w}_2\right)\biggr]\Dx z.
 \end{aligned}
\end{equation}
Finally, the work done by the axial load $P$ is given by:
\begin{equation}
  P\mathcal{E} = \frac{P}{2} \int_0^L \left[ q_s^2 \pi^2 \cos^2 \frac{\pi
      z}{L} - \left( \dot{u}_1 + \dot{u}_2 \right)+ 2\Delta \right] \Dx z,
\end{equation}
where $\mathcal{E}$ comprises the longitudinal displacement due to
global buckling, the in-plane displacement due to local buckling and
the initial end shortening. Note that the displacement due to local
buckling is taken as the average value between the maximum in-plane
displacement in the more compressed outstand $u_1$ and the maximum
in-plane displacement in the less compressed outstand $u_2$, which is
illustrated in Figure \ref{fig:flange_uw}. Moreover, the possible term
in $q_{s0}$ has been neglected since it would vanish on
differentiation for equilibrium anyway. The total potential energy $V$
is therefore assembled thus:
\begin{equation}
   V = U_{bo} + U_{bl} + U_m - P\mathcal{E}.
\end{equation}

\subsection{Variational Formulation}

The governing differential equations are obtained by performing the
calculus of variations on the total potential energy $V$ following a
well established procedure that has been detailed in
\cite{HW1998}. The integrand of the total potential energy $V$ can be
expressed as the Lagrangian ($\mathcal{L}$) of the form:
\begin{equation}
  V = \int_0^L \mathcal{L}\left(\ddot w_i, \dot{w}_i, w_i, \dot {u}_i,
    u_i, z\right) \D z. 
\end{equation} 
The first variation of $V$, which is denoted as $\delta V$, is given by:
\begin{equation}
  \delta V=\int_0^L \left[ \frac{\partial \mathcal{L}}{\partial \ddot
      w_i} \delta\ddot w_i + \frac{\partial \mathcal{L}}{\partial \dot w_i}
    \delta\dot w_i + \frac{\partial \mathcal{L}}{\partial w_i} \delta
    w_i + \frac{\partial \mathcal{L}}{\partial \dot u_i} \delta\dot
    u_i + \frac{\partial \mathcal{L}}{\partial u_i} \delta u_i \right] \D z.
\end{equation}
To find the equilibrium states, $V$ must be stationary, which requires
$\delta V$ to vanish for any small change in $w_i$ and $u_i$. By
assuming that $\delta\ddot w_i=\Dx(\delta\dot w_i)/\D z$, $\delta\dot
w_i=\Dx (\delta w_i)/\Dx z$ and similarly $\delta \dot u_i = \Dx
(\delta u_i)/\Dx z$, integration by parts allows the development of
the Euler--Lagrange equations for $w_i$ and $u_i$; these comprise
fourth order ordinary differential equations (ODEs) in terms of $w_i$
and second order ODEs in terms for $u_i$. For the equations to be
solved by the continuation package \textsc{Auto}, the system variables
need to be rescaled with respect to the non-dimensional spatial
coordinate $\tilde z = 2z/L$. Non-dimensional out-of-plane
displacements $\tilde{w}_i$ and in-plane displacements $\tilde{u}_i$
are also introduced as $2w_i/L$ and $2u_i/L$ respectively. Note that
these scalings assume symmetry about the midspan and the differential
equations are solved for half the length of the strut; this assumption
has been shown to be perfectly acceptable for cases where the global
buckling is critical \cite{Wadee2000}. For cases where local buckling
is critical, this condition is also acceptable so long as the length
of the strut $L$ is much larger than the flange outstand width $b/2$;
hence the critical loads for symmetric and antisymmetric modes are
sufficiently close for the buckling plate. The non-dimensional
differential equations for $w_i$ and $u_i$ are thus:
\begin{equation}
 \begin{split}
   \tilde{\ddddot{w_i}} &- 6\phi^2 \left(1-\nu \right)
   \tilde{\ddot{w}}_i - (-1)^{i} \left(\frac{3\tilde{D}}{8} \right)
   \biggl\{ \left(q_t-q_{t0}\right) \frac{\pi^2}{4\phi} \left(\sin
     \frac{\pi \tilde{z}}{2}
     \tilde{\ddot{w}}_i + \frac{\pi}{2} \cos \frac{\pi
       \tilde{z}}{2} \tilde{\dot{w}}_i \right) \\
   & - (-1)^{i} \biggl[ \tilde{\ddot{w}}_i \left( \frac23 \Delta -
     \frac35 \tilde{\dot{w}}_i^2 \right) - \frac12 \left(
     \tilde{\ddot{u}}_i \tilde{\dot{w}}_i + \tilde{\dot{u}}_i
     \tilde{\ddot{w}}_i \right) \biggr] \biggr\} \\
   & - \frac{3\tilde{G}}{8} \phi^2 \tilde{w}_i \biggl[ \frac23
   {\tilde{\dot{w}}_i^2} + \frac23 {\tilde{w}_i} \tilde{\ddot{w}}_i +
   \tilde{\dot{u}}_i - (-1)^{i} \left(q_s - q_t-q_{s0}+q_{t0}\right)
   \frac{\pi^2}{2\phi} \sin
   \frac{\pi \tilde{z}}{2} \biggr] = 0,\\
 \end{split}
\end{equation}
\begin{equation}
 \begin{split}
   \tilde{\ddot{u}}_i & + \frac34 \tilde{\dot{w}}_i \tilde{\ddot{w}}_i
   + (-1)^{i} \biggl\{ \left(q_t-q_{t0}\right) \frac{\pi^3}{4\phi}
   \cos \frac{\pi \tilde{z}}{2} \\
   & - \left(\frac{3\tilde{G}\phi^2}{\tilde{D}} \right) \biggl[
     \left(q_s - q_t-q_{s0}+q_{t0}\right)\frac{\pi}{\phi} \cos \frac{\pi
       \tilde{z}}{2}
      + (-1)^{i}\left( \frac12 \tilde{w}_i \tilde{\dot{w}}_i
     + \tilde{u}_i \right) \biggr] \biggl\}= 0, \\
   \end{split}
\end{equation}
where $i=\{1,2\}$ again along with $\tilde{D}=EtL^2/D$,
$\tilde{G}=GtL^2/D$ and $\phi=L/b$. Equilibrium also requires the
minimization of the total potential energy with respect to the
generalized coordinates $q_s$, $q_t$ and $\Delta$. This essentially
provides three integral conditions, in non-dimensional form:
\begin{equation}
\begin{split}
  \frac{\partial V}{\partial q_s} &=\pi^2
  \left(q_s-q_{s0}\right)+\tilde{s}\left(q_s -
    q_t-q_{s0}+q_{t0}\right)-\frac{PL^2}{EI_w}q_s\\
  & \quad -\frac{\tilde{s}\phi}{\pi}\int_0^1
  \cos \frac{\pi\tilde{z}}{2} \left[\frac12
    \left(\tilde{w}_1 \tilde{\dot{w}}_1-\tilde{w}_2 \tilde{\dot{w}}_2
    \right)
    +\left(\tilde{u}_1-\tilde{u}_2 \right)\right] \Dx \tilde{z} =0,\\
  \frac{\partial V}{\partial q_t} &= \pi^2 \left(q_t-q_{t0}\right) -
  \tilde{t}\left(q_s - q_t-q_{s0}+q_{t0}\right)+\phi \int_0^1 \biggl\{
  \frac{\tilde{t}}{\pi}
  \cos \frac{\pi\tilde{z}}{2} \biggl[\frac12
    \left(\tilde{w}_1 \tilde{\dot{w}}_1-\tilde{w}_2
      \tilde{\dot{w}}_2\right) \\
  & \qquad 
  +\left(\tilde{u}_1 - \tilde{u}_2\right) \biggr]
  - \sin \frac{\pi \tilde{z}}{2} \left[
    2\left(\tilde{\dot{u}}_1 - \tilde{\dot{u}}_2\right)+\frac34
    \left({\tilde{\dot{w}}_1^2}-{\tilde{\dot{w}}_2^2} \right) \right]
  \biggr\} \Dx \tilde{z} =0,\\
  \frac{\partial V}{\partial \Delta} & = \int_0^1
  \left[ 2\left(1+\frac{h}{b}\right)\Delta -\frac12
  \left(\tilde{\dot{u}}_1+\tilde{\dot{u}}_2\right) - \frac16
  \left(\tilde{\ddot{w}}_1^2 + \tilde{\ddot{w}}_2^2 \right)
  -\frac{P}{Etb} \right] \Dx
  \tilde{z} =0,
\end{split}
\label{eq:equil_int}
\end{equation}
where $\tilde{s}=2GtbL^2/(EI_w)$ and $\tilde{t}=12G\phi^2/E$. Since
the strut is an integral member, the expressions in Equation
(\ref{eq:equil_int}) provide a relationship linking $q_s$ and $q_t$
before any interactive buckling occurs, \emph{i.e.}\ when
$w_i=u_i=0$. This relationship is assumed to hold also between
$q_{s0}$ and $q_{t0}$, which has the beneficial effect of reducing the
number of imperfection amplitude parameters to one. The relationship
between $q_{s0}$ and $q_{t0}$ is given by:
\begin{equation}
q_{s0}=\left(\frac{\pi^2}{\tilde{t}}+1\right)q_{t0}.
\end{equation}

The boundary conditions for $\tilde{w}_i$ and $\tilde{u}_i$ and their
derivatives are for pinned end conditions for $\tilde{x}=0$ and for
symmetry at $\tilde{x}=1$:
\begin{equation}
  \label{eq:bc_w}
  \tilde{w}_i(0) = \tilde{\ddot{w}}_i(0) = \tilde{\dot{w}}_i(1)
  = \tilde{\dddot{w}}_i(1) = \tilde{u}_i(1) = 0,
\end{equation}
with further conditions from matching the in-plane strain:
\begin{equation}
  \label{eq:bc_ud}
  \frac13 \tilde{\dot{u}}_i(0) + \frac18 \tilde{\dot{w}}_i^2(0) - \frac12
  \Delta + \frac{P}{2Etb} = 0.
\end{equation}

Linear eigenvalue analysis for the perfect strut ($q_{s0}=q_{t0}=0$)
is conducted to determine the critical load for global buckling
$\Pco$. This is achieved by considering that the Hessian matrix
$\mathbf{V}_{st}$ at the critical load is singular. Hence:
\begin{equation}
  \mathrm{det} \left( \mathbf{V}_{st} \right) = \left|
    \begin{array}{cc}
      \frac{\partial^2 V}{\partial q_s^2} & \frac{\partial^2
        V}{\partial q_s \partial q_t} \\
      \frac{\partial^2 V}{\partial q_t \partial q_s} & \frac{\partial^2
        V}{\partial q_t^2}
    \end{array}
  \right| = 0,
\end{equation}
Recalling of course that in fundamental equilibrium for this case,
$q_s = q_t = w_i = u_i = 0$. Hence, the critical load for global
buckling is:
\begin{equation}
  \Pco=\frac{\pi^2 EI_w}{L^2}+\frac{2Gtb}{1+\tilde{t}/\pi^2}.
\label{eq:pc}
\end{equation}
If the limit $G \rightarrow \infty$ is taken, which represents a
principal assumption in Euler--Bernoulli bending theory, the critical
load expression converges to the Euler buckling load for an I-section
strut buckling about the weak axis.

\section{Numerical examples of perfect behaviour}

The full nonlinear differential equations are obviously complicated to
be solved analytically. The continuation and bifurcation software
\textsc{Auto-07p} \cite{Auto2007} has been shown in the literature
\cite{WG2012,HW1998} to be an ideal tool to solve the equations
numerically. For this type of mechanical problem, one of its major
attributes is that it has the capability to show the evolution of the
solutions to the equations with parametric changes. The solver is very
powerful in locating bifurcation points and tracing branching paths as
model parameters are varied. To demonstrate this, an example set of
cross-section and material properties are chosen which are shown in
Table \ref{tab:Secprop}.
\begin{table}[htb]
\centering
 \begin{tabular}{rl}
   \hline
   Flange width $b$ & $96~\mm$ \\ 
   Flange thickness $t$ & $1.2~\mm$ \\
   Cross-section depth $h$ & $120~\mm$ \\
   Cross-section area $A$ & $513~\mm^2$\\
   Young's modulus $E$ & $210~\mathrm{kN/mm^2}$ \\
   Poisson's ratio $\nu$ & 0.3\\
   \hline
 \end{tabular}
 \caption{Cross-section and material properties of an example
   strut. Recall that the thickness of the web $t_w = 2t$. The
   geometric properties are similar to those tested in
   \protect\cite{Becque2009expt}. The length $L$ is varied such that
   the cases where global buckling or local buckling are critical can
   be presented.}
 \label{tab:Secprop}
\end{table}
In this example, perfect behaviour is assumed and hence $W_0 =
\theta_0 =0$. The global critical load $\Pco$ can be calculated using
Equation (\ref{eq:pc}), whereas an estimate for the local buckling
critical stress $\sigma_l^\mathrm{C}$ can be evaluated using the
well-known plate buckling formula
$\sigma_l^\mathrm{C}=kD\pi^2/(b^2t)$, where the coefficient $k$
depends on the boundary conditions; approximate values of $k=0.426$
and $k=4$ are chosen for the rectangular plates representing the
flange outstands (three edges pinned and one edge free) and the web
(all four edges pinned) respectively, assuming that the plates are
relatively long \cite{Bulson}. Table \ref{tab:plpc}
\begin{table}[htb]
\centering
\begin{tabular}{ccccc}
  $L~(\mathrm{m})$ & $\sigma_o^\mathrm{C}~(\mathrm{N/mm^2})$ &
  $\sigma_{l,\mathrm{flange}}^\mathrm{C}~(\mathrm{N/mm^2})$ &
  $\sigma_{l,\mathrm{web}}^\mathrm{C}~(\mathrm{N/mm^2})$ &
  Critical mode\\
  \hline
  $3.5$ & $58.3 $ & $51.1$ & $2731$ & Local (flange)\\
  $4.0$ & $44.7$ & $51.1$ & $2731$ & Global\\
  \hline
\end{tabular}
\caption{Theoretical values of the global and local critical buckling
  stresses for two separate lengths. The expression for
  $\sigma_o^\mathrm{C}=\Pco/A$ and the web is obviously not vulnerable
  to local buckling.}
\label{tab:plpc}
\end{table}
summarizes the critical stresses and shows that the assigned
cross-section dimensions satisfy the assumptions that the local mode
is critical for one of the lengths and the global mode is critical for
the other. Moreover, the critical stress of the web is orders of
magnitude higher than that of the flange, which justifies the
assumption stated earlier. It should be emphasized that the local
buckling critical stress is calculated numerically in \textsc{Auto}
and is usually marginally higher than the value given in Table
\ref{tab:plpc}, with an error not exceeding $5\%$ with the theoretical
expressions given above for the long plates.

\begin{figure}[htbp]
  \centering
  \subfigure[]{\psfig{figure=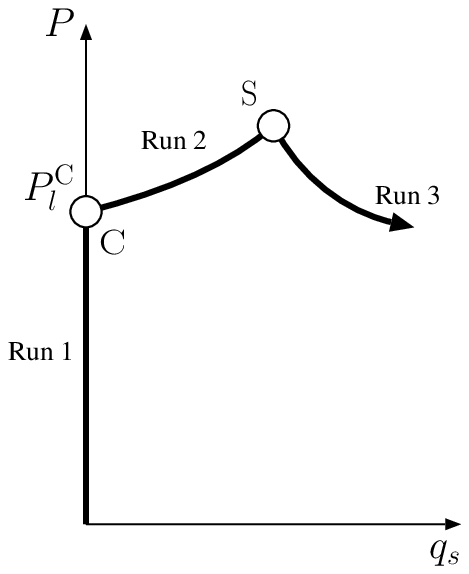,width=46mm}}\quad
  \subfigure[]{\psfig{figure=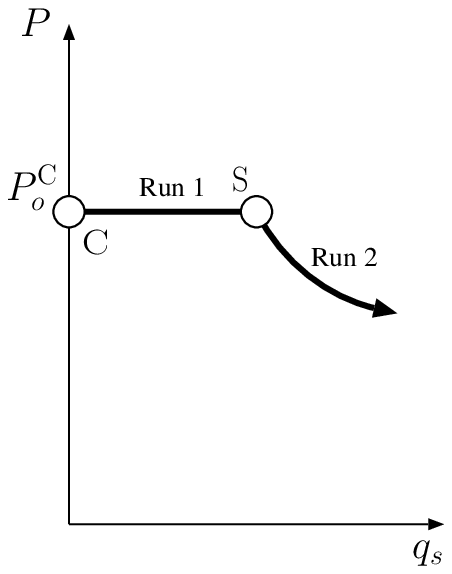,width=46mm}}\quad
  \subfigure[]{\psfig{figure=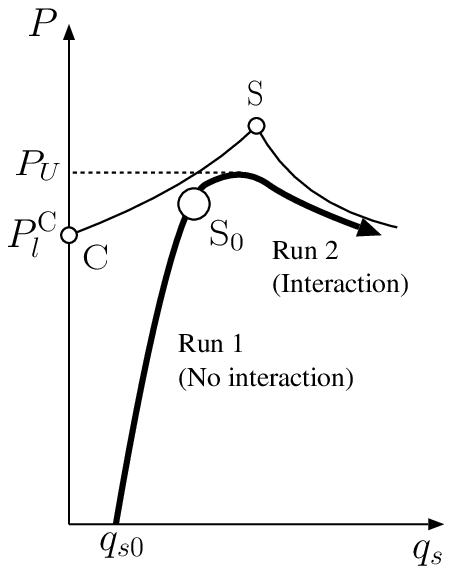,width=46mm}}
  \caption{Numerical continuation procedures. (a) Local buckling being
    critical; (b) Global buckling being critical; (c) Imperfect case
    -- example of local buckling being critical shown. The thicker
    line shows the actual solution path in each of the examples shown.
    Points $\mathrm{C}$ and $\mathrm{S}$ represent the critical and
    secondary bifurcations respectively, whereas the point
    $\mathrm{S}_0$ represents the bifurcation leading to interactive
    buckling in the imperfect case with the load $P_U$ being the
    ultimate load in the imperfect case.}
  \label{fig:autoruns}
\end{figure}

Numerical continuation was performed in \textsc{Auto} for the cases
where local buckling and global buckling were critical in turn. The
principal parameters used in the continuation process were
interchangeable, but generally $q_s$ was varied for computing the
equilibrium paths for the distinct buckling modes and $P$ was varied
for evaluating the interactive buckling paths. For the case of local
buckling being critical, the continuation process initiated from zero
load with the local buckling critical load $\Pcl$ being obtained
numerically. The post-buckling path was then computed by using the
branch switching facility within the software and the distinct local
buckling equilibrium path was computed until a secondary bifurcation
point $\mathrm{S}$ was found. It was from this point that the
interactive buckling path was found, again through the use of branch
switching.  For the case where global buckling was critical, since the
critical load was determined analytically in Equation (\ref{eq:pc}),
the initial post-buckling path was computed first from $\Pco$ and many
bifurcation points were detected on the weakly stable post-buckling
path; the focus being on the one with the lowest value of $q_s$, the
secondary bifurcation point $\mathrm{S}$. A subsequent run was then
necessary starting from $\mathrm{S}$ using the branch switching
function, after which the equilibrium path again exhibits the
interaction between the global and the local modes.  Figure
\ref{fig:autoruns} shows the procedures for the cases diagrammatically
with (a--b) concerning the perfect cases discussed above and (c), the
imperfect case, is considered later in the section on validation.

\subsection{Local buckling critical}

\begin{figure}[htb]
\centering
\subfigure[]{\includegraphics[scale=0.78]{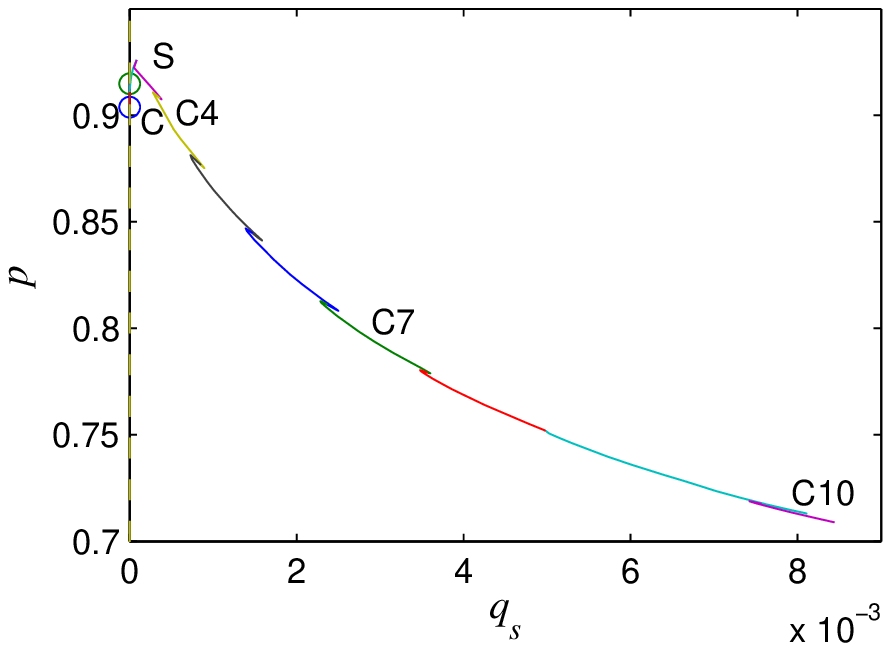}}
\subfigure[]{\includegraphics[scale=0.78]{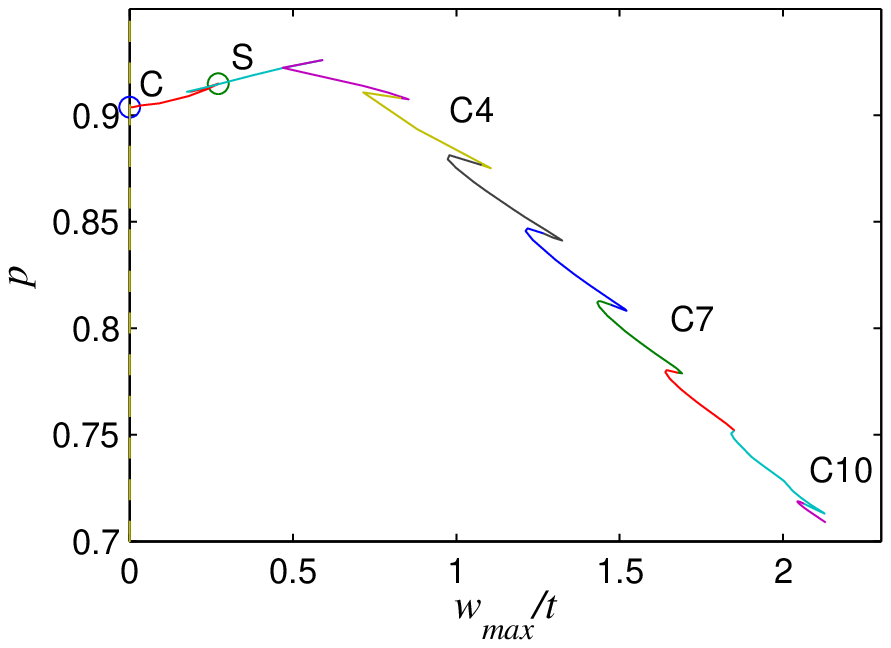}}
\subfigure[]{\includegraphics[scale=0.78]{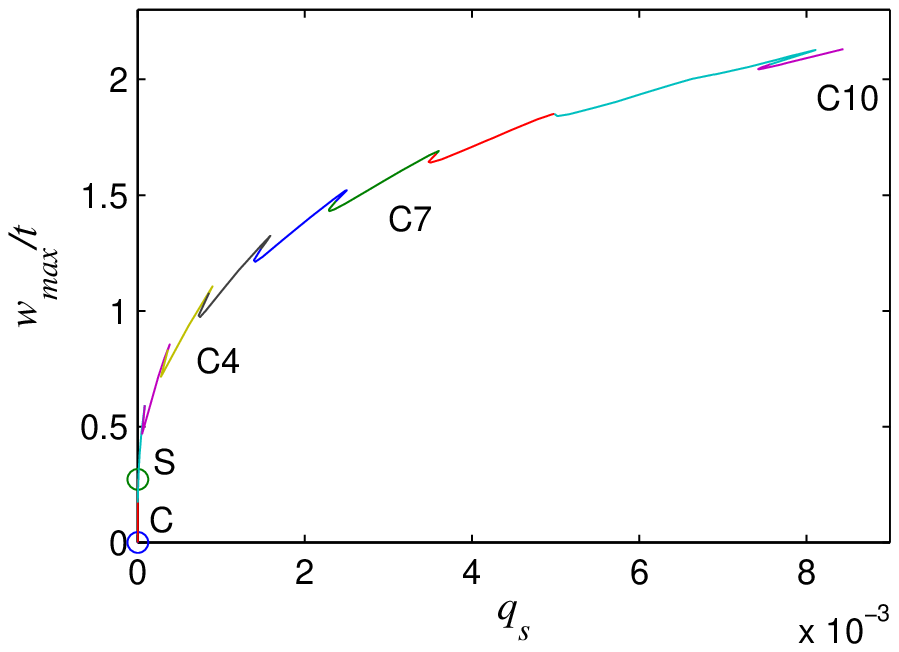}}
\subfigure[]{\includegraphics[scale=0.78]{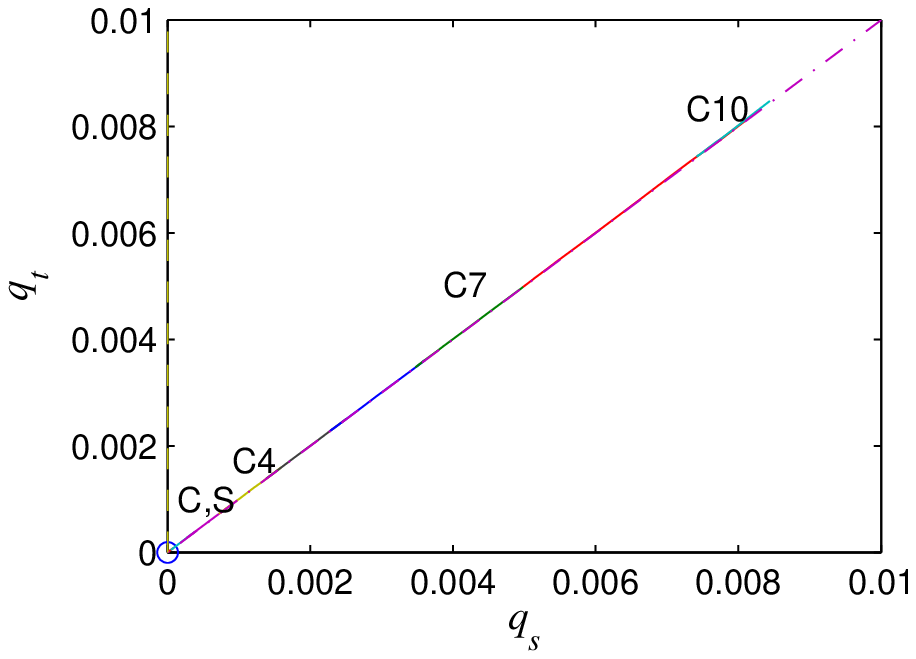}}
\caption{Numerical equilibrium paths for $L=3.5~\mathrm{m}$ where
  local buckling is critical. Graphs of the normalized force ratio $p$
  versus (a) the generalized coordinate $q_s$ and (b) the maximum
  out-of-plane displacement of the buckled flange plate, in
  non-dimensional form, $w_\mathrm{max}/t$ are shown. (c) shows
  $w_\mathrm{max}/t$ versus $q_s$ and (d) shows the relationship
  between the generalized coordinates $q_s$ and $q_t$ defining the
  global buckling mode during interactive buckling, with the dot-dash
  line showing the Euler--Bernoulli bending condition $q_s = q_t$.}
\label{fig:equil_local}
\end{figure}

In this section, the strut with properties given in Table
\ref{tab:Secprop} with length $L$ being $3.5~\mathrm{m}$ is analysed,
where the flanges buckle locally first. Figure \ref{fig:equil_local}
shows a plot of the normalized axial load $p=P/\Pco$ versus (a) the
global mode and (b) the local mode amplitudes; (c) shows the local and
global mode relative magnitudes during post-buckling and (d) shows
that there is a small but importantly, non-zero shear strain during
global buckling. The local critical buckling load is calculated at
$p=0.905$, whereas according to Table \ref{tab:plpc}, this value
should be $0.877$, which represents a small error of $3\%$,
particularly since it is well known that the theoretical expression
for the critical buckling stress for the long plate is usually an
underestimated value.

\begin{figure}[htb]
\centering
\includegraphics[scale=0.7]{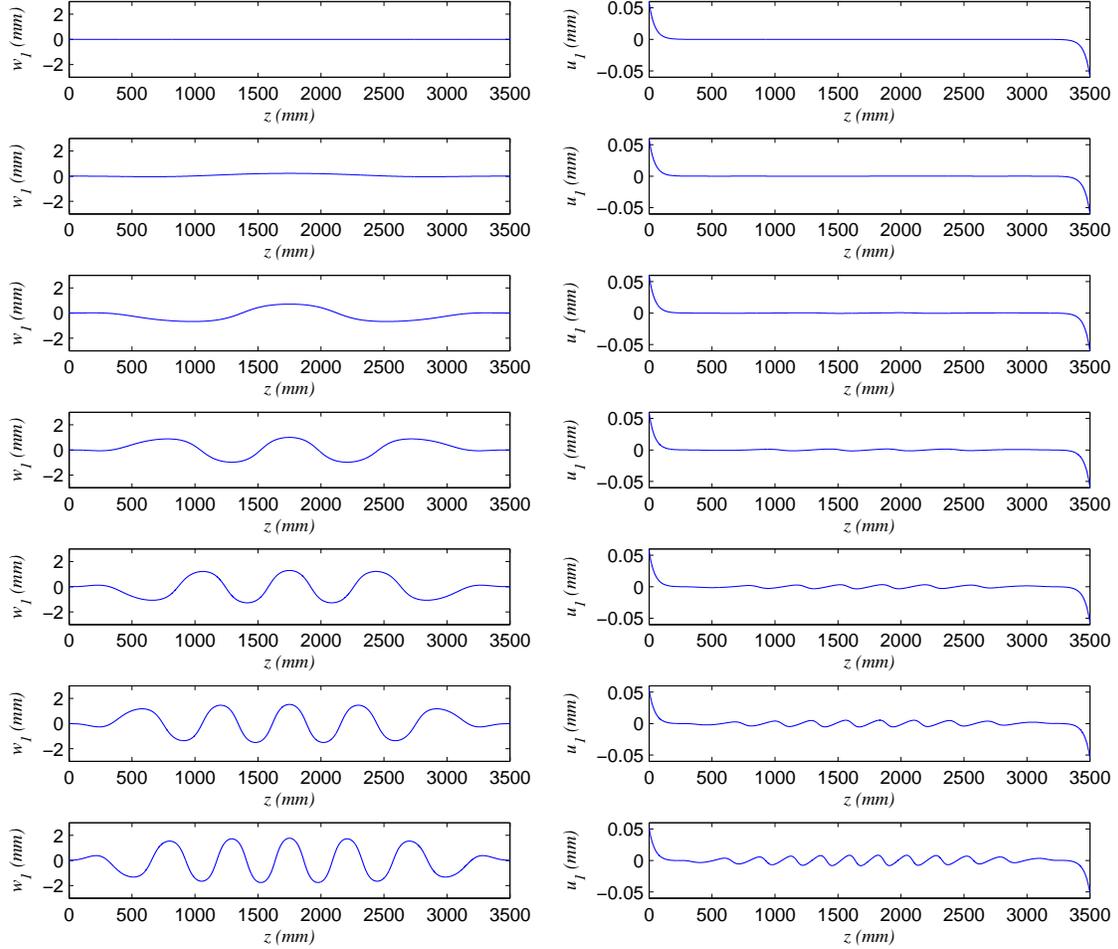}
\caption{Numerical solutions for the local out-of-plane displacement $w_1$
  (left) and local in-plane displacement $u_1$ (right) for the tip
  ($x=-b/2$) of the vulnerable flange. Individual solutions on
  equilibrium paths $C1$ to $C7$ are shown in sequence from top
  to bottom respectively.}
\label{fig:w1u1_local}
\end{figure}

\begin{figure}[htbp]
  \centering
  \subfigure[]{\includegraphics[scale=0.37]{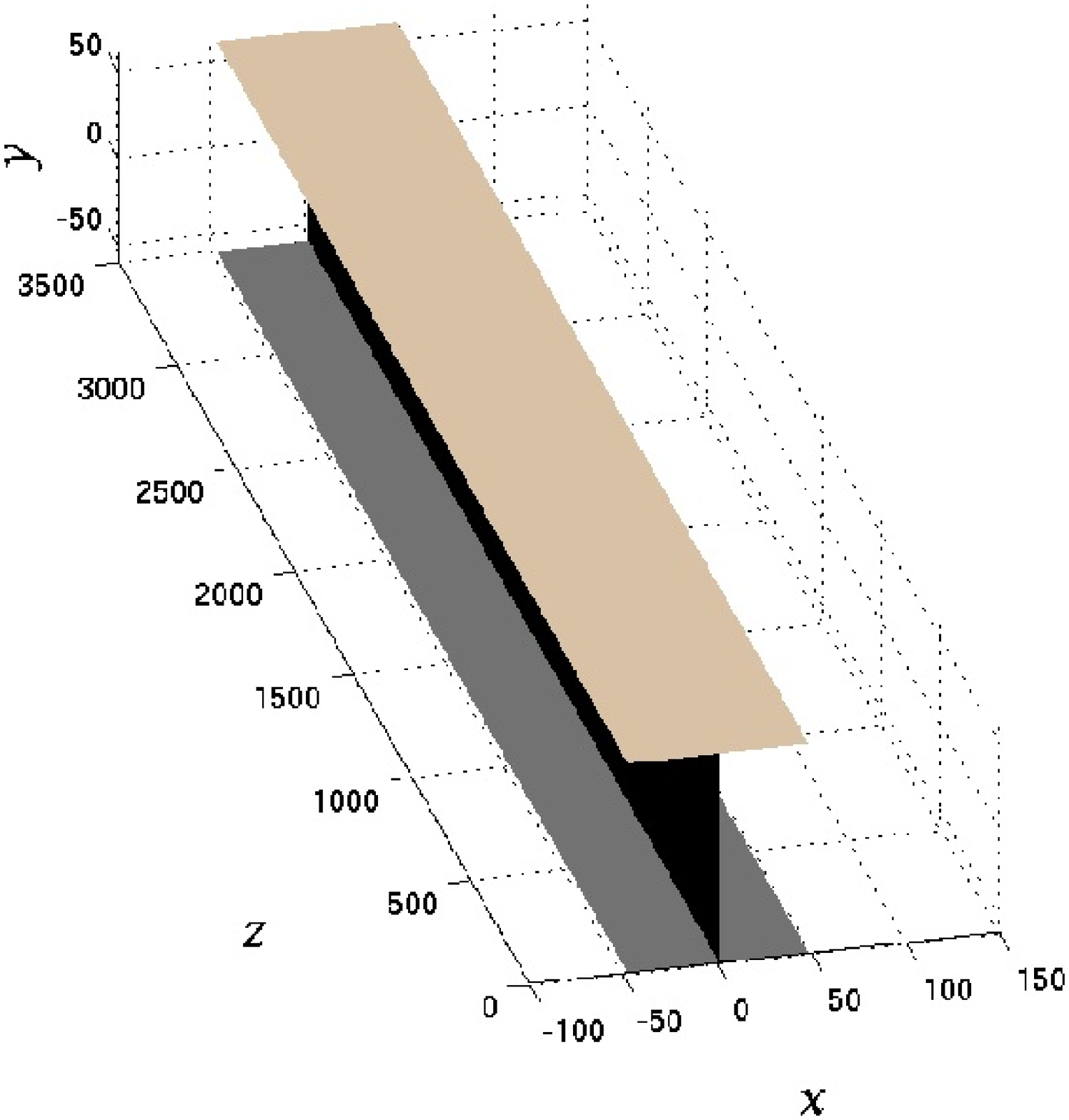}}
  \subfigure[]{\includegraphics[scale=0.37]{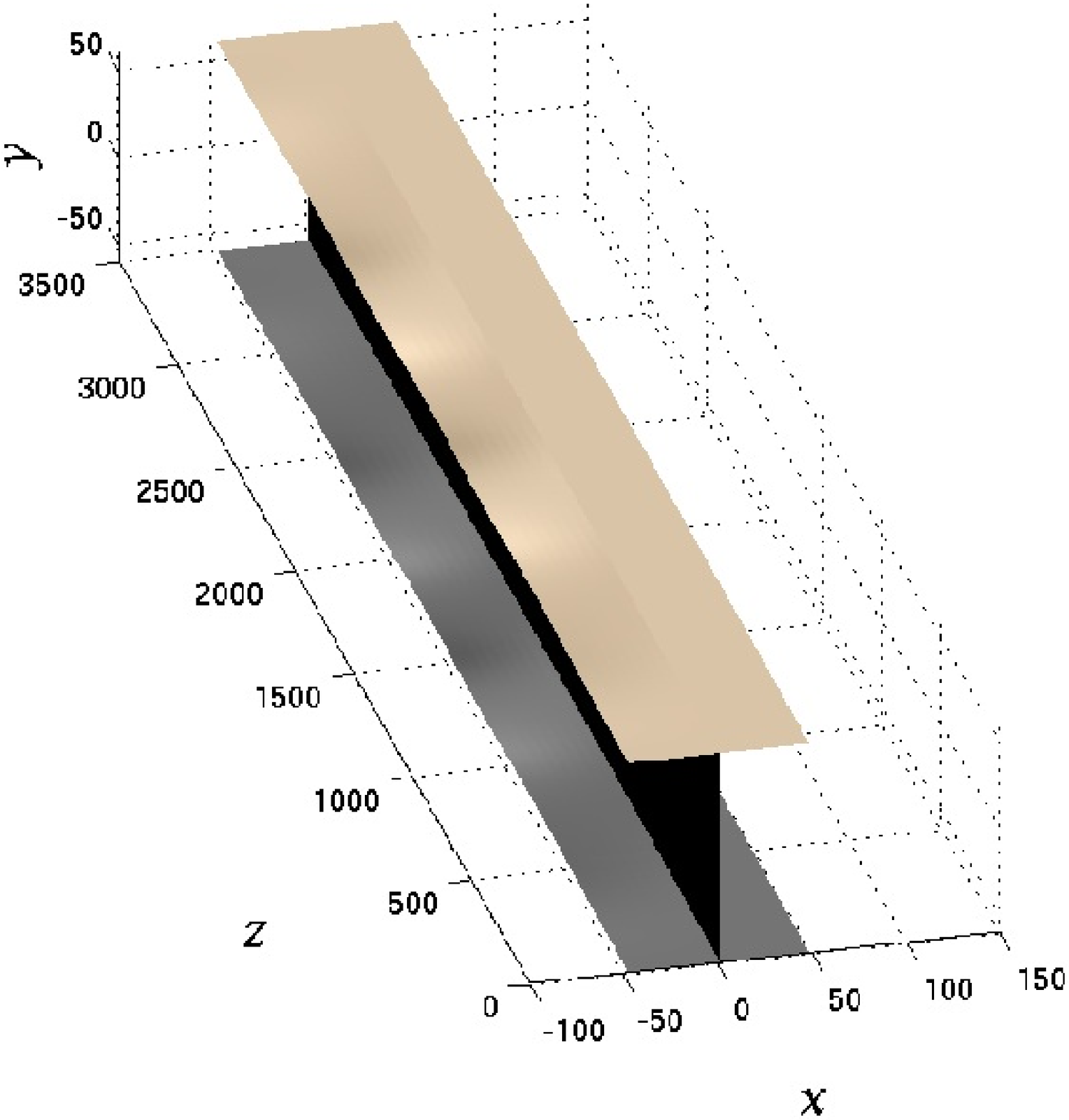}}
  \subfigure[]{\includegraphics[scale=0.37]{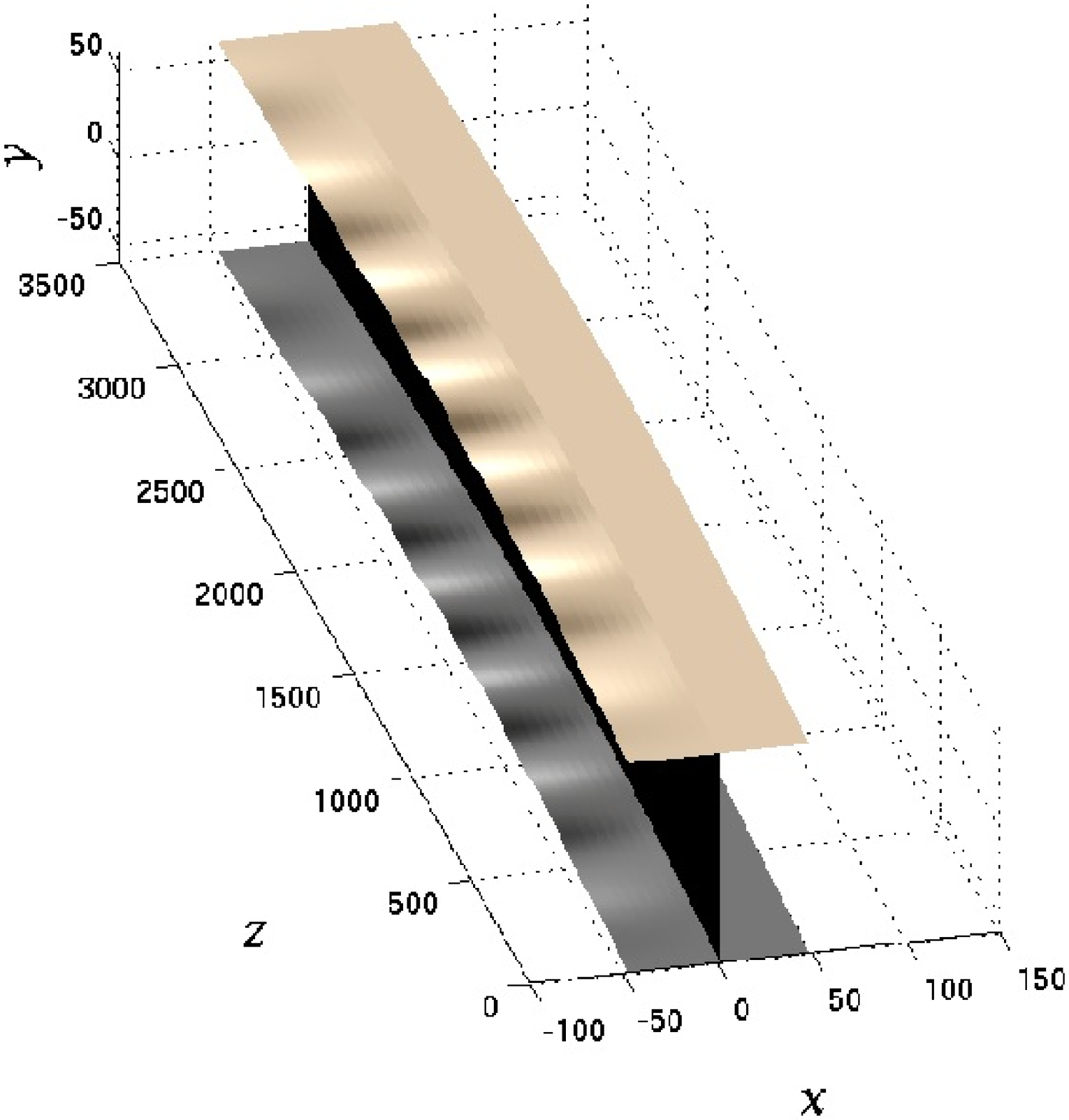}}
  \subfigure[]{\includegraphics[scale=0.37]{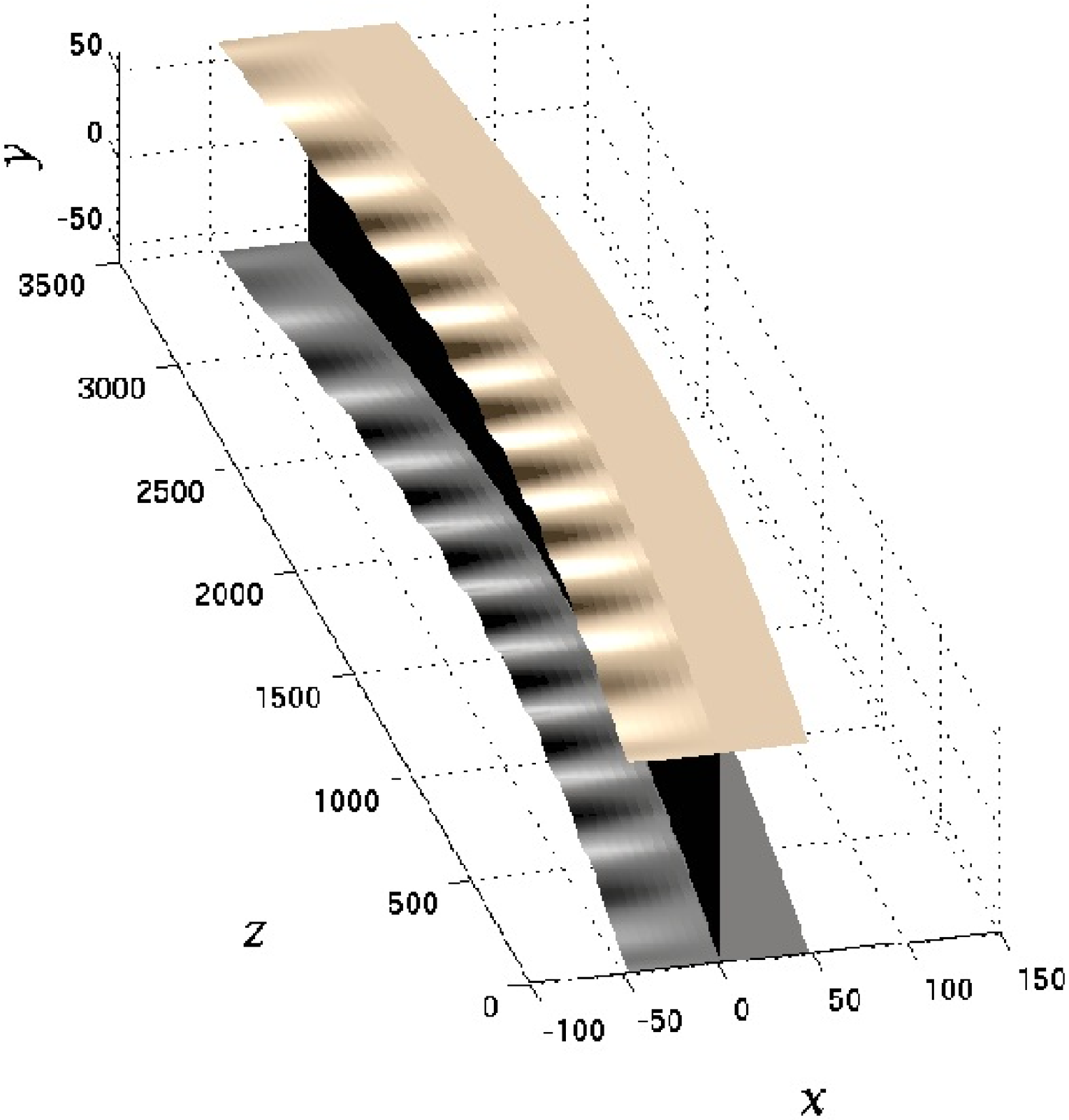}}
  \caption{Numerical solutions of the system of equilibrium equations
    visualized on 3-dimensional representations of the strut. The
    results are shown for individual points on paths (a)
    $C1~(p=0.9039)$, (b) $C4~(p=0.9081)$, (c) $C7~(p=0.8111)$ and (d)
    $C10~(p=0.7177)$. All dimensions are in millimetres.}
\label{fig:cells_local}
\end{figure}

\begin{figure}[htb]
\centering
\includegraphics[scale=0.7]{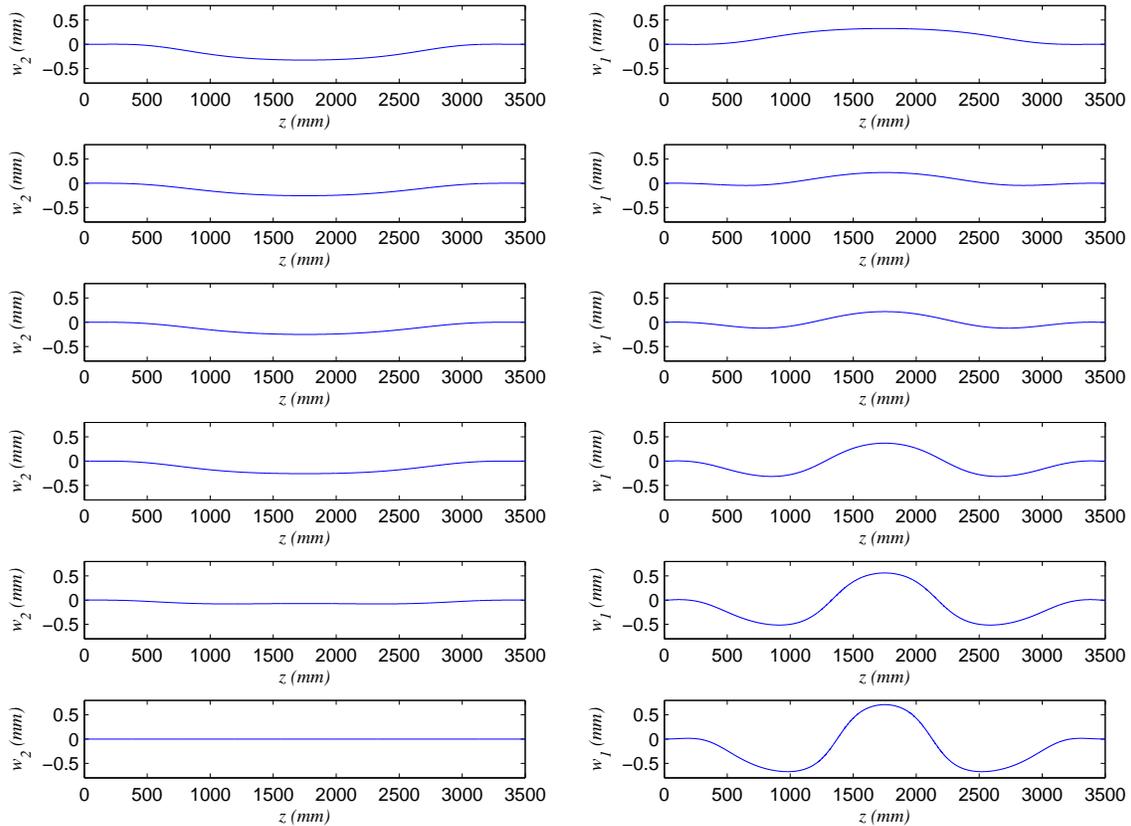}
\caption{Numerical solutions for the local out-of-plane displacement
  $w_2$ (left) for the tip of the non-vulnerable flange ($x=b/2$) and
  $w_1$ (right) for the tip of the vulnerable flange ($x=-b/2$) for
  cells 1--3. Note the rapid decay of $w_2$ reflecting the reducing
  compression in that outstand once global buckling is triggered.}
\label{fig:w2w1}
\end{figure}

One of the most distinctive features of the equilibrium paths, as
shown in Figures \ref{fig:equil_local}(a)--(c), is the sequence of
snap-backs that effectively separates the equilibrium path into $10$
individual parts (or \emph{cells}) in total as shown. The fourth,
seventh and the tenth paths are labelled as $C4$, $C7$ and $C10$
respectively. Each path or cell corresponds to the formation of a new
local buckling displacement peak or trough. Figure
\ref{fig:w1u1_local}
illustrates the corresponding progression of the numerical solutions
for the local buckling functions $w_1$ and $u_1$ from cell $C1$ to
$C7$, where $C1$ represents the initial post-buckling equilibrium path
generated from $\mathrm{C}$. Once a secondary bifurcation is triggered
at $\mathrm{S}$, it is observed that the local buckling mode is
contaminated by the global mode and interactive buckling ensues with
the buckling deformation spreading towards the supports as new peaks
and troughs are formed. Figure \ref{fig:cells_local}
shows a selection of 3-dimensional representations of the deflected
strut that comprise the components of global buckling ($W$ and
$\theta$) and local buckling ($w_i$ and $u_i$) at a specific state on
paths $C1$, $C4$, $C7$ and $C10$. As the equilibrium path develops to
$C10$, the maximum out-of-plane displacement $w_\mathrm{max}$ approaches
a value of $2.5~\mm$ which is roughly twice the flange thickness and
can be regarded as large in terms of geometric assumptions. The
interactive buckling pattern becomes effectively periodic on path
$C10$. Any further deformation along the equilibrium path would be
expected to cause restabilization to the system since the boundaries
would begin to confine the spread of the buckling deformation. It
should be stressed of course that any plastic deformation during the
loading stage would destabilize the system significantly. Figure
\ref{fig:w2w1}
shows the comparison between the lateral displacement of the two
flange outstands. The local buckling displacement in the
non-vulnerable outstand $w_2$ decays to zero rapidly as the global
mode amplitude increases during interactive buckling; by the third
cell, $w_2$ has vanished implying that if global buckling occurs
first, both $w_2$ and $u_2$ would be negligible.

The magnitude of direct and shear strains may be calculated once the
governing differential equations are solved. The direct strain in the
non-vulnerable part of the flange becomes tensile at $C10$ due to
bending, whereas the maximum direct strain in the vulnerable part of
the flange is approximately $1.3\e{-3} (=0.13\%)$. This level of
strain is confined to the ends of the strut and is also well below the
yield strain of most structural steels; moreover for the stainless
steels given in Becque and Rasmussen \cite{Becque2009expt,Becque_thesis},
significant strain softening only begins from approximately $0.15\%$
strain and so quantitative comparisons can be made for the
post-buckling response for the majority of the cells.

Systems that exhibit the phenomenon described above are termed in the
literature to show ``cellular buckling'' \cite{Hunt2000} or
``snaking'' \cite{BurkeKnobloch2007}. In such systems, progressive
destabilization and restabilization is exhibited; currently, the
destabilization is caused primarily by the interaction of the global
and local instabilities, whereas the restabilization is caused by the
stretching of the buckled plates when they bend into double
curvature. As the amplitude of the global buckling mode $q_s$
increases, the compressive bending stress in the flange outstands
increase also, which imply that progressively longer parts of the
flange are susceptible to local buckling. Since local buckling is
inherently stable, the drop in the load from the unstable mode
interaction is limited due to the stretching of the plate when it
buckles into progressively smaller wavelengths. Therefore, the
cellular buckling occurs due to the complementary effects of the
unstable mode interaction and stable local buckling.

\subsection{Global buckling critical}

\begin{figure}[htb]
\centering
\subfigure[]{\includegraphics[scale=0.78]{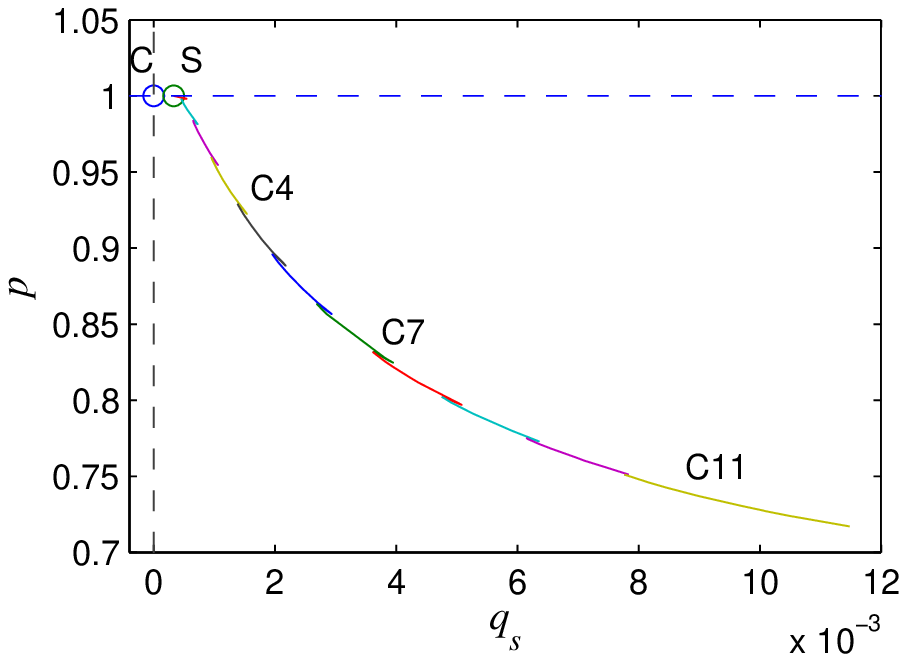}}
\subfigure[]{\includegraphics[scale=0.78]{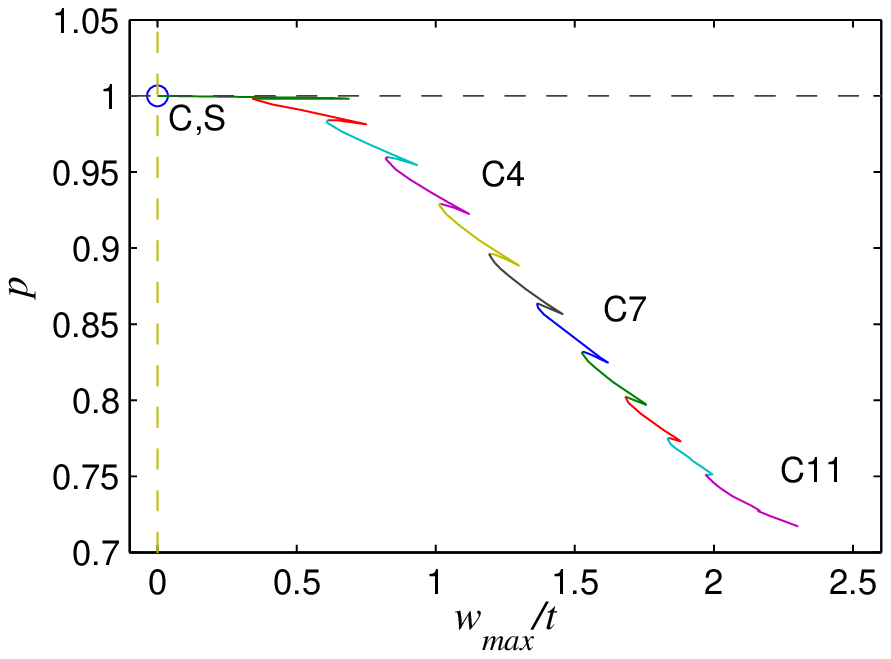}}
\subfigure[]{\includegraphics[scale=0.78]{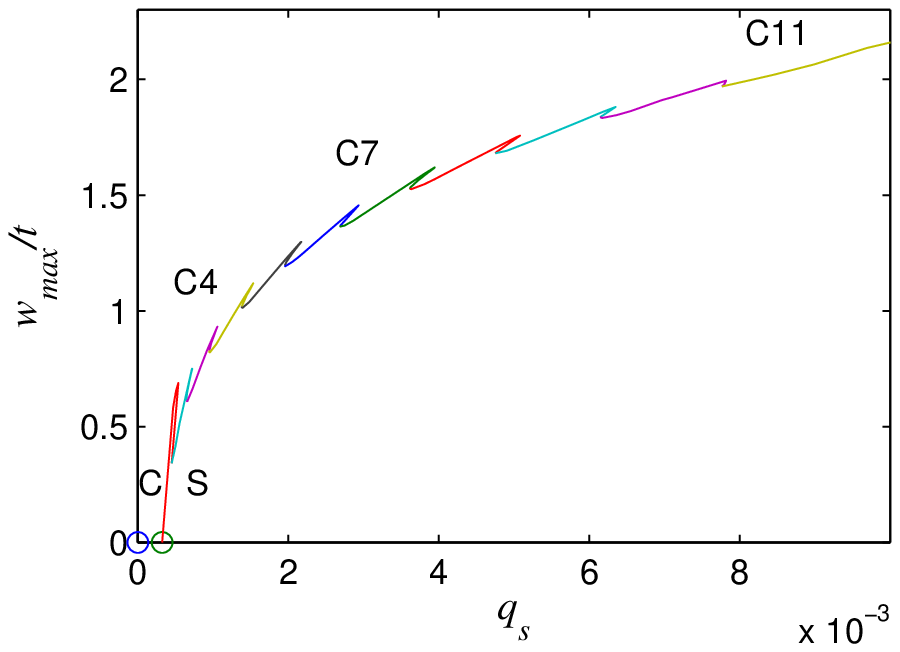}}
\subfigure[]{\includegraphics[scale=0.78]{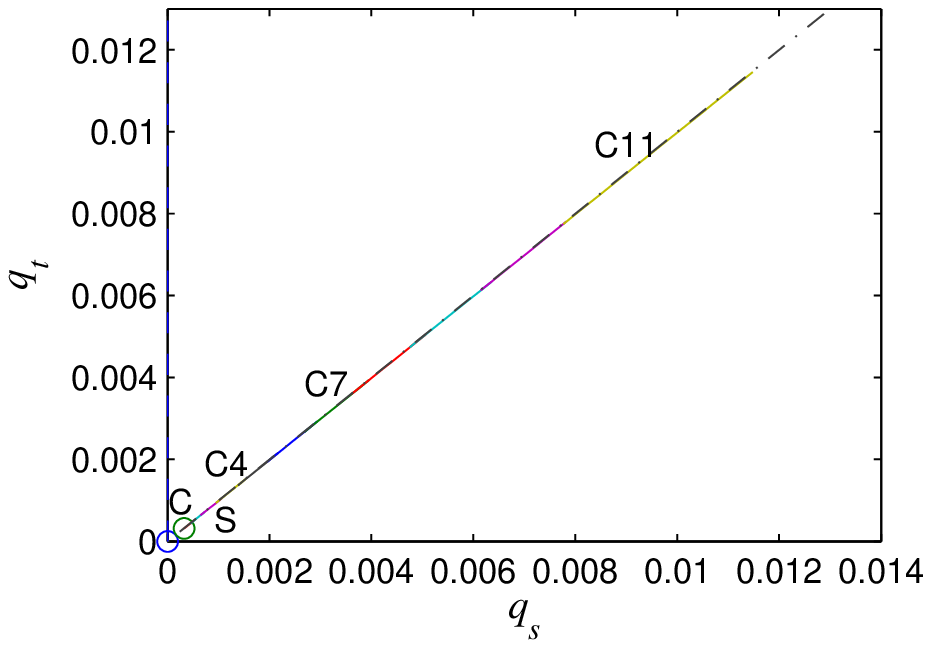}}
\caption{Numerical equilibrium paths for $L=4.0~\mathrm{m}$ where
  global buckling is critical. Graphs of the normalized force ratio
  $p$ versus (a) the generalized coordinate $q_s$ and (b) the maximum
  out-of-plane displacement of the buckled flange plate, in
  non-dimensional form $w_\mathrm{max}/t$, are shown. (c) shows
  $w_\mathrm{max}/t$ versus $q_s$ and (d) shows the relationship
  between the generalized coordinates $q_s$ and $q_t$ defining the
  global buckling mode during interactive buckling, with the dot-dash
  line showing the Euler--Bernoulli bending condition $q_s = q_t$.}
\label{fig:equil_global}
\end{figure}

The strut with properties given in Table \ref{tab:Secprop} with length
$L$ being $4.0~\mathrm{m}$ is now analysed; in this case, global Euler
buckling occurs first. Figure \ref{fig:equil_global}
shows plots of the equilibrium diagrams that correspond directly to
Figure \ref{fig:equil_local}. This time, cellular buckling is
triggered when the pure global mode is contaminated by the local
mode. Since the global mode is only weakly stable, no significant
post-buckling stiffness is exhibited initially. Moreover, since the
global mode places the non-vulnerable flange outstand into less
compression before any local buckling occurs, the functions $w_2$ and
$u_2$ can be neglected as a consequence of the observations made in
connection with Figure \ref{fig:w2w1}; this simplifies the formulation
considerably.

The emergence of the buckling cells in sequence is very similar to
that shown for the case where local buckling is critical and so it is
not presented in detail for brevity. Nevertheless, with the model in
place, quantitative comparisons can be made against existing
experiments.

\section{Validation and discussion}

\subsection{Comparison with experiments of Becque and Rasmussen}

A recent experimental study of thin-walled I-section struts by Becque
and Rasmussen \cite{Becque2009expt,Becque_thesis} focused on the case
where local buckling is critical. Although the struts were made from a
stainless steel alloy (ferritic AISI404), the compressive
stress--strain curve showed that the material remained linearly
elastic when the strain was below approximately $0.15\%$. Two specific
tests were conducted on struts with material and geometric properties
as given in Table \ref{tab:becprop}.
\begin{table}[htb]
\centering
 \begin{tabular}{rll}
   \hline
   Strut length $L$ & $3.0~\mathrm{m}$ & $2.5~\mathrm{m}$\\
   Flange width $b$ & $96.64~\mm$ & $96.80~\mm$\\ 
   Corner Radius $r$ & $3.06~\mm$ & $3.02~\mm$\\
   Flange thickness $t$ & $1.21~\mm$ & $1.21~\mm$\\
   Section depth $h$ & $125.12~\mm$ & $125.24~\mm$\\
   \hline
 \end{tabular}
 \caption{Geometric properties for the strut tests taken directly from
   \protect\cite{Becque2009expt,Becque_thesis}. Recall that the
   thickness of the web $t_w = 2t$. For both struts the initial
   Young's modulus $E = 195~\mathrm{kN/mm^2}$ and Poisson's
   ratio $\nu = 0.3$.}
 \label{tab:becprop}
\end{table}
The initial out-of-straightness mid-length lateral deflections of the
specimens of length $L$ being $3~\mathrm{m}$ and $2.5~\mathrm{m}$ were
measured to be $L/3352$ and $L/16234$ respectively
\cite{Becque_thesis}. In order to make direct comparisons, numerical
runs were conducted in \textsc{Auto} with the initial global buckling
mode imperfection amplitude ratio $q_{s0}$ being equal to $3 \times
10^{-4}$ and $6 \times 10^{-5}$ respectively. The cross-section
properties given in Table \ref{tab:becprop} were adapted slightly to
consider the effective width of the flange, $b_e = b-2r$; the
effective width was used in the numerics for the analytical model, the
results of which follow.

\begin{figure}[htb]
\centering
\subfigure[]{\includegraphics[scale=0.78]{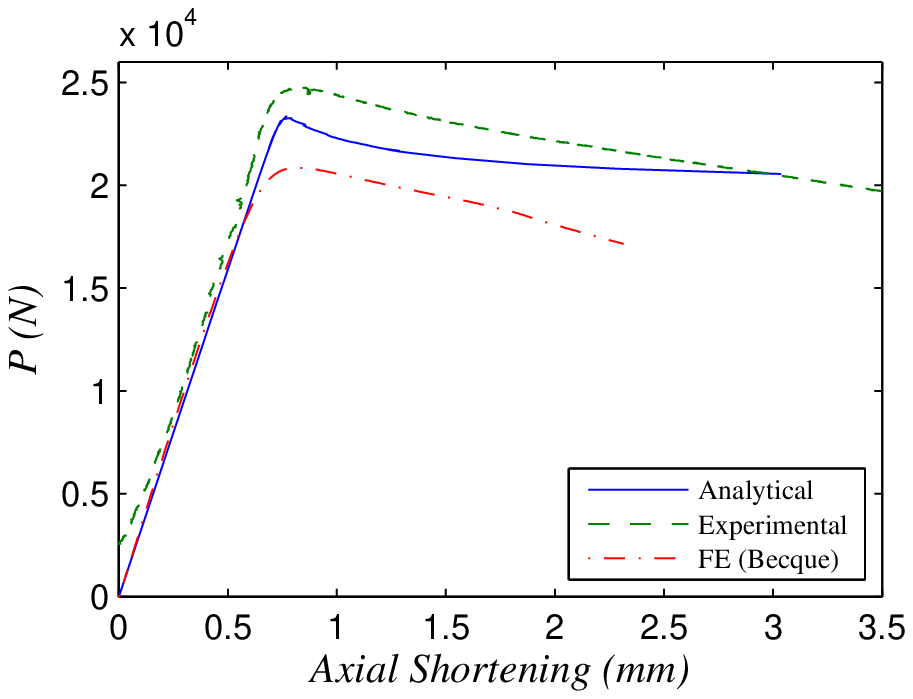}}
\subfigure[]{\includegraphics[scale=0.78]{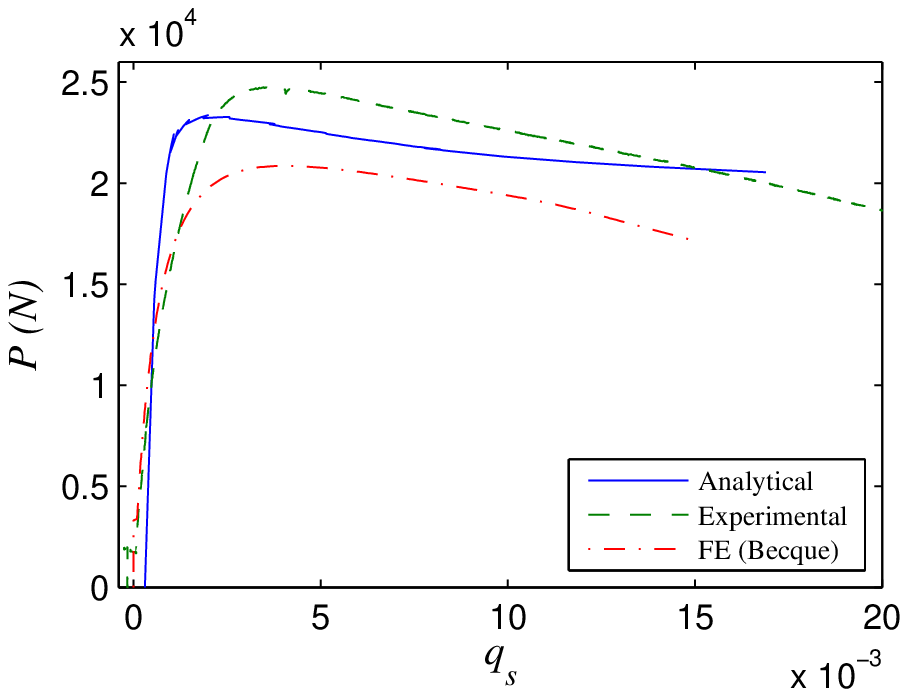}}
\subfigure[]{\includegraphics[scale=0.78]{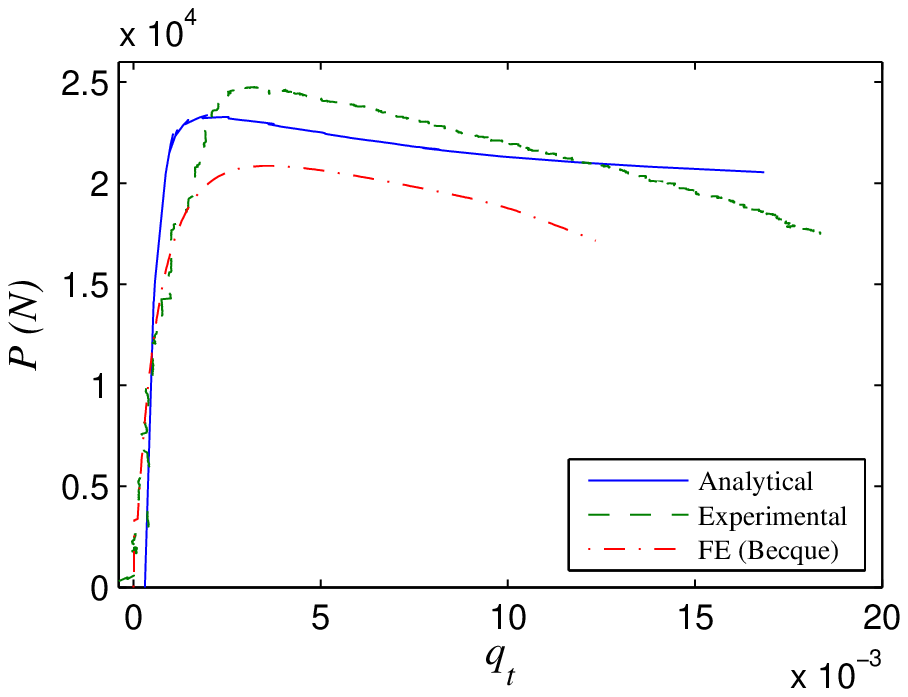}}
\subfigure[]{\includegraphics[scale=0.78]{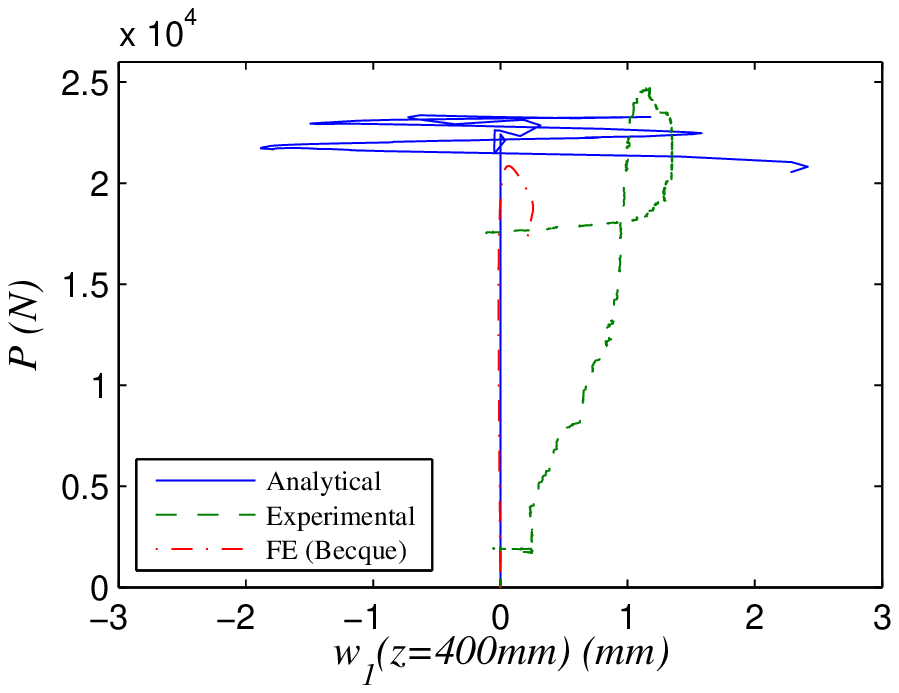}}
\caption{Numerical equilibrium paths comparing with Becque's
  experiment with a strut of length $3~\mathrm{m}$. Graphs of the
  applied axial load $P$ versus (a) the total end shortening, (b) the
  generalized coordinate $q_s$, (c) the generalized coordinate $q_t$,
  (d) the out-of-plane displacement of the buckled flange plate $w_1$
  measured at $z=400~\mm$ are shown.  Solid lines show the current
  analytical model, whereas the dashed and dot-dashed lines
  respectively show the experimental and finite element results from
  \protect\cite{Becque_thesis}.}
\label{fig:equil_valid_3000}
\end{figure}

\begin{figure}[htb]
\centering
\subfigure[]{\includegraphics[scale=0.78]{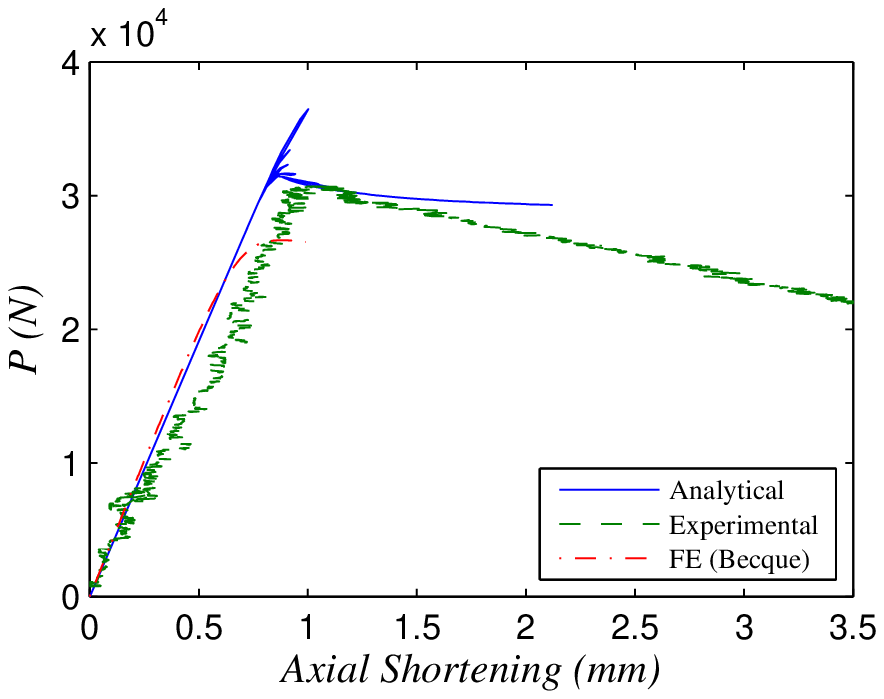}}
\subfigure[]{\includegraphics[scale=0.78]{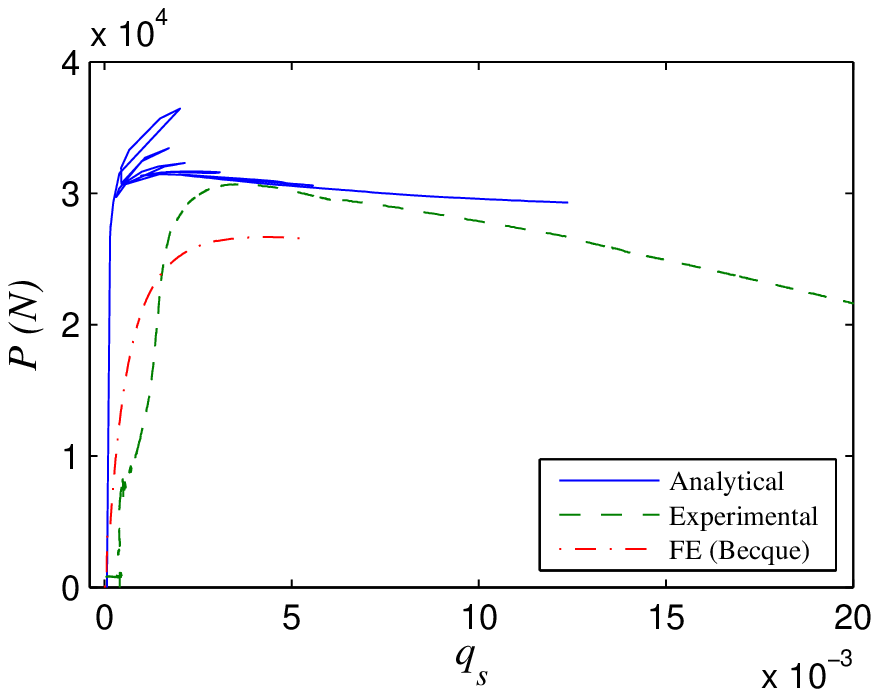}}
\subfigure[]{\includegraphics[scale=0.78]{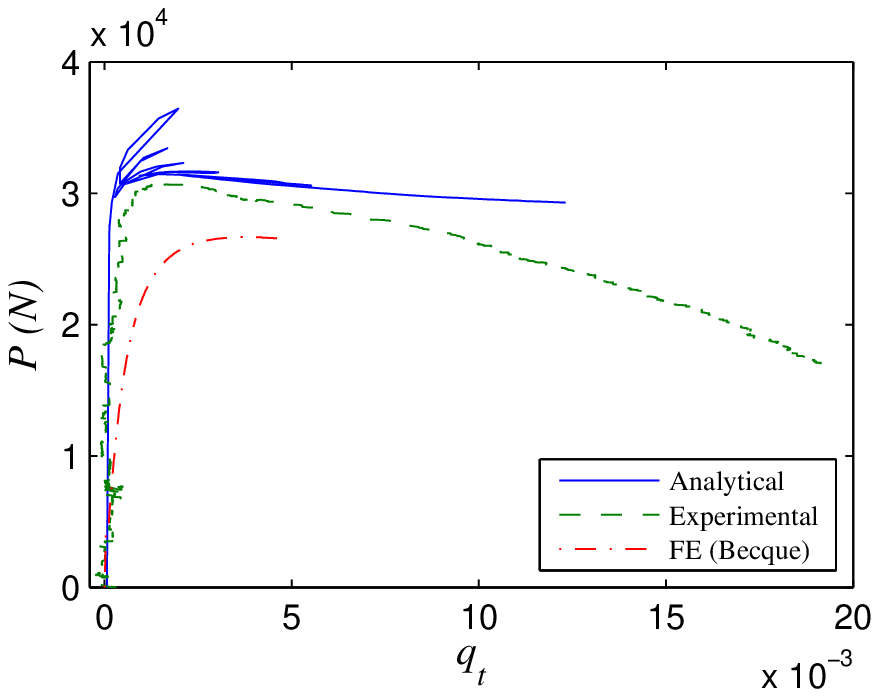}}
\subfigure[]{\includegraphics[scale=0.78]{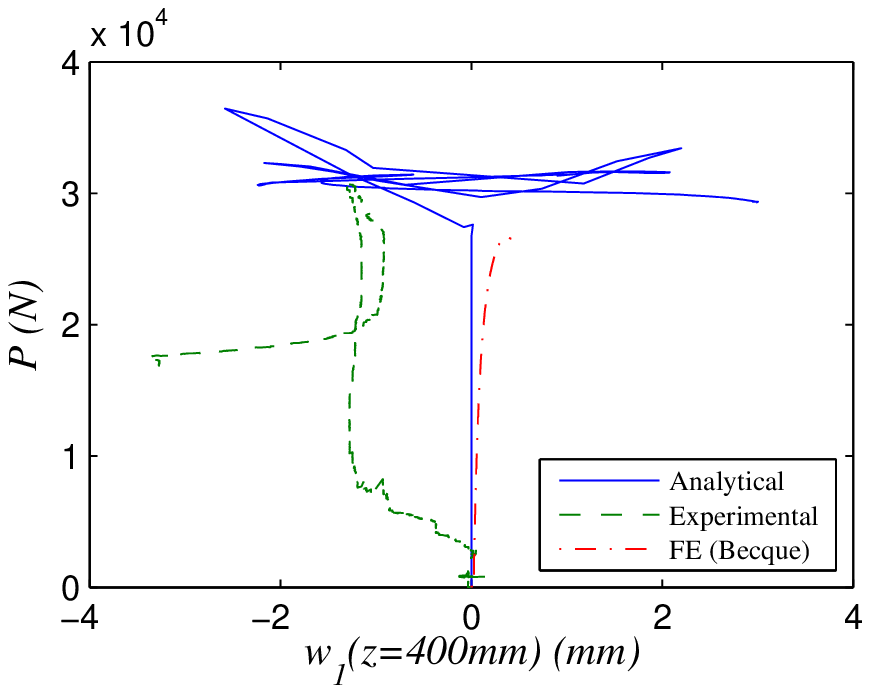}}
\caption{Numerical equilibrium paths comparing with Becque's
  experiment with a strut of length $2.5~\mathrm{m}$. Graphs of the
  applied axial load $P$ versus (a) the total end shortening, (b) the
  generalized coordinate $q_s$, (c) the generalized coordinate $q_t$,
  (d) the out-of-plane displacement of the buckled flange plate $w_1$
  measured at $z=400~\mm$ are shown. Solid lines show the current
  analytical model, whereas the dashed and dot-dashed lines
  respectively show the experimental and finite element results from
  \protect\cite{Becque_thesis}.}
\label{fig:equil_valid_2500}
\end{figure}

The numerical continuation process was initiated from zero load with
the process being illustrated in Figure \ref{fig:autoruns}(c). The
value of $q_s$ was increased up to a bifurcation point, shown as
$\mathrm{S}_0$ in Figure \ref{fig:autoruns}(c), after which
interactive buckling was introduced. The equilibrium path then
progressed to a limit point at which $P$ can be defined as the
ultimate load $P_U$. Then destabilization and the cellular buckling
behaviour was observed as described in the previous section. Figures
\ref{fig:equil_valid_3000}
and \ref{fig:equil_valid_2500}
show comparisons between the current analytical model, the
experimental results from \cite{Becque2009expt,Becque_thesis} and the
numerical models from \cite{Becque2009num,Becque_thesis}.  The
comparisons show strong agreement between the analytical model and the
results from the physical experiments, the correlation being clearly
superior to the previous numerical results presented in
\cite{Becque2009num,Becque_thesis}.

\begin{figure}[htbp]
  \centerline{\psfig{figure=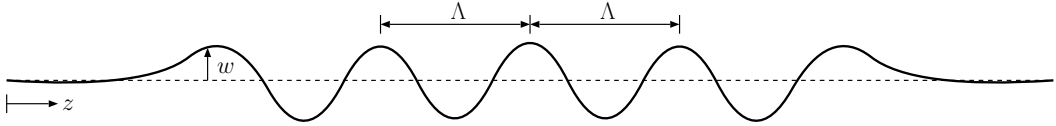,width=140mm}}
  \caption{Definition of local buckling wavelength $\Lambda$ from
    results for $w (\equiv w_1)$ from the variational model.}
  \label{fig:waves}
\end{figure}

For the $3~\mathrm{m}$ length strut, the ultimate load was found to be
$25.2~\kN$ from the experiment, which is approximately $3\%$ higher
than the numerical value from the analytical model where
$P_U=24.4~\kN$. It is also observed in Figure
\ref{fig:equil_valid_3000}(d) that the theoretical local out-of-plane
displacement $w_1$, at a location that was remote from the strut
midspan ($z=400~\mm$), changes from positive to negative and
\emph{vice versa} several times. This is a signature of the cellular
behaviour, indicating the progressive change in wavelength of the
local buckling mode pattern. The actual experimental response, on the
other hand, is always likely to jump to the final cell relatively
rapidly once the initial instability is triggered. This was shown in
the experiments conducted during work on the interactive buckling of
beams \cite{WG2012}, particularly in the cases where global and local
buckling were triggered at similar load levels. The reason for this is
that in a physical experiment, even with displacement control, the
mechanical response in the region with snap-backs exhibits dynamic,
rather than static, behaviour. Although in the current case the
experiment did not pick up the full cellular response, it did show the
change from positive to negative for $w_1$, which is a clear
indication of the changing wavelength in the local buckling mode
pattern. The interactive buckling wavelength $\Lambda$ can also be
compared, which is defined in Figure \ref{fig:waves}.
The local buckling mode had a plate buckling wavelength that was
measured to be $275~\mm$ with a modulated amplitude for this specific
test \cite{Becque2009expt}. The numerical results in the current work
show that the value of $\Lambda$ is $280~\mm$ for the interactive
buckling wavelength at the end of the equilibrium paths from the
analytical model shown in Figure \ref{fig:equil_valid_3000}. The close
comparison (less than $2\%$ difference) offers further grounds for
encouragement for future developments of the current model.

For the $2.5~\mathrm{m}$ length strut, the features are similar; the
ultimate load $P_U$ is $16\%$ higher than the maximum load shown in
the experiment. However, this is only a very small part of the global
picture. The graphs in Figure \ref{fig:equil_valid_2500}(a--c) show
sequential snap-backs in the theoretical response almost immediately
in the post-buckling range that reduce the true load-carrying capacity
to levels which practically coincide with the experimental result,
demonstrating an excellent overall comparison. In Figure
\ref{fig:equil_valid_2500}(d), a similar response is observed to
Figure \ref{fig:equil_valid_3000}(d). The change in sign of the
out-of-plane displacement from the test clearly demonstrates the
changing buckling wavelength again. Unfortunately, a numerical
measurement of the plate buckling wavelength was not reported for this
particular experiment.

Finally, it can be seen that the results from the analytical model and
the experiments begin to diverge after a certain level of
displacement. This is postulated to be as a result of material softening
due to the use of stainless steel in the experiments, whereas linear
elasticity is assumed throughout the analytical model. However, for
the most part the close comparisons between the analytical model and
the experimental results, in the authors' opinion, validate the
current modelling approach both qualitatively and quantitatively.

\subsection{Future model enhancements}

The success in capturing the interactive buckling behaviour allows for
some speculation of how the current work may be extended. The
technical difficulty of capturing sharp successive snap-back
instabilities numerically most probably explains why the previous
finite element models \cite{Becque2009num}, although giving safe
predictions for the global strength of the tested struts in
\cite{Becque2009expt,Becque_thesis}, showed a relatively indifferent
comparison with the experiments. A possible way around this problem in
future finite element modelling of such struts could be to introduce
in turn a sequence of initial imperfections with different shapes that
resemble the modes from each cell. This would allow an envelope of the
nonlinear equilibrium solutions to be computed that resemble the
actual post-buckling response.

Other issues that can be investigated are those regarding the
assumption of the fixity between the elements of the cross-section,
namely between the web and the flanges, and the introduction of lipped
ends to the flanges. In terms of joint fixity, the flange--web
junctions are modelled as pinned and hence are free to rotate. By
modelling them as partially to fully rigid, a more extensive range of
responses could be captured. With the resultant increase in structural
stiffness in the cross-section this would introduce to the system, the
local buckling load would definitely increase. However, the early
evidence from a pilot study is that the post-buckling becomes less
cellular as a result \cite{SEMC2013}. A similar effect may be obtained
by attaching or designing flanges with lips to reduce the
vulnerability to local buckling. However, lips introduce the
possibility of distortional buckling \cite{Schafer2002} which, in this
context, is known to resemble localized buckling \cite{HW91,WHW97}
rather than the cellular buckling found presently. Localization, in
this context, can be more severely destabilizing than the cellular
buckling. Work on this latter enhancement is currently in the early
stages and hence the point regarding the potentially greater severity
in the post-buckling instability is purely conjecture currently.

\section{Concluding remarks}

A nonlinear analytical model based on variational principles has been
presented for axially-loaded thin-walled I-section struts buckling
about the weak axis of bending. The model identifies an important and
potentially dangerous interaction between global and local modes of
instability, which leads to highly unstable cellular buckling through
a series of snap-back instabilities. These result from the increasing
contributions of buckling mode amplitudes forcing the flanges in more
compression to buckle progressively. This process had also been
observed in recent experimental work and in other components that
suffer from a nonlinear interaction between global and local
buckling. Comparisons with published experiments are excellent and
validate the model. Extending the analytical approach would allow
further study of the parameters that drive the behaviour and provide a
greater and more profound understanding of the underlying
phenomena. This, in turn, would provide designers with the information
about the sensitivity of thin-walled components to small changes in
geometry.

\section*{Acknowledgement}

The authors would like to thank Dr Jurgen Becque of the University of
Sheffield, UK, for technical discussions and allowing us to use his
experimental results.

\bibliography{refs}

\end{document}